\documentclass[twocolumn]{aastex63}

\submitjournal{AJ}

\shorttitle{How to detect the limbs of exoplanets}
\shortauthors{Espinoza \& Jones}
\definecolor{ForestGreen}{HTML}{228B22}
\usepackage{amsmath}

\begin{document}

\title{Constraining mornings \& evenings on distant worlds: a new semi-analytical approach and prospects with transmission spectroscopy}

\correspondingauthor{N\'estor Espinoza}
\email{nespinoza@stsci.edu}

\author[0000-0001-9513-1449]{N\'estor Espinoza}
\affiliation{Space Telescope Science Institute, 3700 San Martin Drive, Baltimore, MD 21218, USA}

\author[0000-0002-2316-6850]{Kathryn Jones}
\affiliation{University of Bern, Center for Space and Habitability, Gesellschaftsstrasse 6, CH-3012, Bern, Switzerland}
%\affiliation{Max-Planck-Institut f\"ur Astronomie, K\"onigstuhl 17, 69117 Heidelberg, Germany}

\received{}
\revised{}
%\accepted{}

%\collaboration{1}{(AAS Journals Data Scientists collaboration)}




\begin{abstract}

The technique of transmission spectroscopy --- the variation of the planetary radius with wavelength due to opacity sources in the planet’s terminator region --- has been to date one of the most successful in the characterization of exoplanet atmospheres, providing key insights into the composition and structure of these distant worlds. {{A common assumption made 
when using this technique}}, however, is that the variations are the same in the entire terminator region. In reality, the morning and evening terminators might have distinct temperature, pressure and thus compositional profiles due to the inherent 3-D nature of the planet which would, in turn, give rise to different spectra on each side of it. Constraining those might be fundamental for our understanding of not only the weather patterns in these distant worlds, but also of planetary formation signatures which might only be possible to extract once these features are well understood. Motivated by this physical picture, in this work we perform a detailed 
study on the observational prospects of detecting this effect. We present an open-source semi-analytical framework with which this information can be extracted directly from transit lightcurves, and perform a detailed study on the 
prospects of detecting the effect with current missions such as \textit{TESS} and upcoming ones such as 
\textit{JWST}. Our results show that these missions show great promise for the detection of this effect. Transmission 
spectroscopy studies with \textit{JWST}, in particular, could provide spectra of each of the limbs allowing us to 
convey 3-D information previously accessible only via phase-curves.

\end{abstract}



\section{Introduction} \label{sec:intro}

The technique of transmission spectroscopy --- the wavelength dependence of the planetary radius during 
transit \citep{ss:2000, hubbard:2001, BSH:2003, fortney:2005}, has been one of the most successful ones 
in the past decade to explore the composition and structure of exoplanet atmospheres, providing key 
insights into their interior structures and compositions \citep[see, e.g.,][for a review]{kreidberg:2018}. 
From an observational perspective, to obtain a transit spectrum researchers typically fit a transit 
model to precise wavelength-dependent lightcurves in order to retrieve the transit depths{{, $(R_p/R_*)^2$,}} as a function 
of wavelength. {{Typically, the fitting procedure relies}} in one simple, 
but key assumption: the terminator region we observe during transit is homogeneous. There is already 
growing evidence that this assumption might actually be unrealistic in relatively hot ($T_{\textnormal{eq}} > 1000$ {{K}}) 
exoplanet atmospheres, where the day-to-night differences might in turn imply different structures and 
overall compositions in their morning and evening\footnote{In this work, the morning and evening limbs are 
also referred to as the leading and trailing limbs, respectively.} terminators \citep[see, e.g.,][and references therein]{fortney:2010, dobbs:2012, kempton:2017, powell:2019,macdonald:2020, helling:2020}. Constraining them 
might give precious insights into circulation patterns and compositional stratification which might probe 
to be fundamental for our understanding of the weather patterns in distant worlds. For example, hazes 
are expected to be photochemically produced and thus they would most likely be able to form in the 
dayside \citep{kempton:2017, powell:2019}. These could, in turn, be transported to the trailing limb, 
while clouds could be transported from the nightside (where they are expected to form due to the 
lower temperatures) into the leading limb, thus resulting in a drastically different transmission 
spectrum between them, and thus effective sizes of the radii of each limb 
\citep{kempton:2017,powell:2019}. Directly detecting this effect would not only serve to put theories like the ones 
proposed by \cite{kempton:2017} and \cite{powell:2019} to the test, but would directly impact on the 
fundamental assumptions of transmission spectroscopy studies to date, implying there is not \textit{one} set of 
properties (e.g., abundances) to extract from transmission spectra. This is, in turn, critical to perform inferences on 
e.g., formation scenarios based on extracted molecular abundances with this technique \citep[see, e.g., ][and references 
therein]{oberg,mordasini, espinoza}.

Previous works \citep[e.g., ][]{dobbs:2012,LP:2016,kempton:2017,powell:2019,macdonald:2020} 
have already studied the prospects and impact of limb inhomogeneities on transit spectra. 
Overall, the consensus seems to be that there is already both observational and theoretical 
evidence that this is an effect that is important to consider and that might even be 
impacting current transit spectra. \cite{LP:2016}, \cite{kempton:2017} and \cite{powell:2019} 
have already laid out the foundation of the theoretical aspects of detecting this effect{{, while 
\cite{macdonald:2020} has in fact studied publicly available transmission spectra in order to 
show that they can actually be explained as arising from two distinct temperature/pressure 
profiles}}. In 
this work, we explore these prospects from an observational perspective which aims at detecting 
the effect of limb asymmetries \textit{directly in transit lightcurves}, such that 
interpretations can be made at a later stage on each of the limbs. 

{{Detecting limb-asymmetries at the lightcurve level could have important impacts 
on how researchers typically approach transit spectroscopy for several reasons. First, it 
could imply that performing inference about the limbs on transit depths obtained through 
lightcurve fits using the classic \cite{ma:2002} symmetric transit model is 
subject to be biased, as the lightcurves would be essentially fit with the wrong model. If 
the limbs have different properties, their transit spectrum -and thus their ``transit depths"- 
would be different, injecting lightcurve asymmetries that these models cannot properly account for. Second, performing inference on the transit depths obtained through these symmetric transit 
models also necessitates a handful of assumptions in order to overcome the degeneracies that 
fitting a single transmission spectrum with two different temperature, pressure and abundance profiles imply. This, in turn, diminishes the discovery space to the assumptions made 
by our models, which might be quite a restrictive imposition, especially in the era of 
ultra-high spectrophotometric precision such as the ones the upcoming \textit{James Webb 
Space Telescope} (\textit{JWST}) will be opening up. Here, we propose instead that 
if indeed limb asymmetries can be detected in the transit lightcurves themselves, this would 
open up a whole new and direct framework for obtaining information about them. In this framework, 
we would be able to extract \textit{two} ``limb spectra" from a given transit lightcurve: 
one spectrum for each limb, which we could interpret individually at a later stage through, 
e.g., atmospheric retrievals and/or forward models. It is important to note that the 
essence of}} this proposition is not particularly new (it has already been suggested by the work of 
\citealt{vParis}). {{Our}} contribution in this work is to perform a deep 
dive into (a) \textit{how} we might actually perform this characterization in a fast and reliable way, 
(b) \textit{what is the level of detectability} of this effect with current and near-future instrumentation 
and (c) to show how, in some cases, this might even be \textit{the most efficient way of extracting this information} from 
transit lightcurves. Some of these points have already been touched upon by \cite{powell:2019} at different degrees of depth; 
here we expand and homogenize the discussion from an observational perspective, which we believe complements these previous 
works on this topic. 

Our work is organized as follows. In Section \ref{sec:thealgorithm} we present a new semi-analytical method to 
extract the transit depths from each of the limbs of an exoplanet. The core idea of this method was actually 
already put forward by \cite{vParis}, where each of the limbs of the exoplanet are modelled as stacked 
semi-circles. However, we expand on this modeling framework in that our calculation is made in a 
semi-analytical fashion, making use of geometrical arguments and the algorithm used by \texttt{batman} 
\citep{batman}. This makes the lightcurve computation much faster than the numerical scheme described in 
\cite{vParis}, and allows us to expand it to account for sky-projected \textit{planetary} spin-orbit misalignments. 
We present a python library to generate lightcurves with this new algorithm, \texttt{catwoman} \citep{jones}, in Section \ref{sec:catwoman}, provide an overview of the model and validate it against a 
numerical implementation in Section \ref{sec:validation}. In Section \ref{sec:detection} we present simulations in 
which we explore the feasibility of detecting this effect with current precise photometry such as that of 
the \textit{Transiting Exoplanet Survey Satellite} mission \citep[\textit{TESS};][]{ricker-tess} and near-future instrumentation such as spectrophotometry to 
be obtained by the upcoming \textit{JWST}. In Section \ref{sec:discussion} 
we present a discussion and implications of our results, along with a case-study on the exoplanet HAT-P-41b, 
which we use to demonstrate how extracting the spectrum of the limbs of this exoplanet might give insights into 
possible models that give rise to the observed transit spectrum by the 
\textit{Hubble Space Telescope} (\textit{HST}). We summarize our main conclusions in Section \ref{sec:conclusions}.

%%%%%%%%%%%%%%%%%%%%%%%%%%%%%%%%%%%%%%%%%%%%%%%%%%
%%%%%%%%%%%%%%%%%%%%%%%%%%%%%%%%%%%%%%%%%%%%%%%%%%
%%%%%%%%%%%%%%%%%%%%%%%%%%%%%%%%%%%%%%%%%%%%%%%%%%
%%%%%%%%%%%%%%%%%%%%%%%%%%%%%%%%%%%%%%%%%%%%%%%%%%
\section{Modeling limb asymmetries in transit lightcurves}
\label{sec:thealgorithm}

The idea proposed by \cite{vParis} to model the signatures of asymmetric limbs in transit lightcurves involves a 
very simple concept: approximate the terminator regions of the leading and trailing limbs as two stacked 
semi-circles with different radii. In essence, the idea is that each limb produces an independent transit spectrum 
that we ought to recover by modeling the lightcurve imprinted by them. In that work, the authors used a numerical 
framework to compute the resulting lightcurve, which is relatively computationally expensive. Here we use the 
same idea but tackle the problem from a different angle: instead of using a numerical approach, we employ a 
semi-analytical framework, which in turn allows for faster lightcurve computations. In this new framework, the stacked semi-circles 
are also allowed to be \textit{rotated} with respect to the orbital motion, expanding thus the proposed framework by 
\cite{vParis}.

\begin{figure*}
\includegraphics[width=2.1\columnwidth]{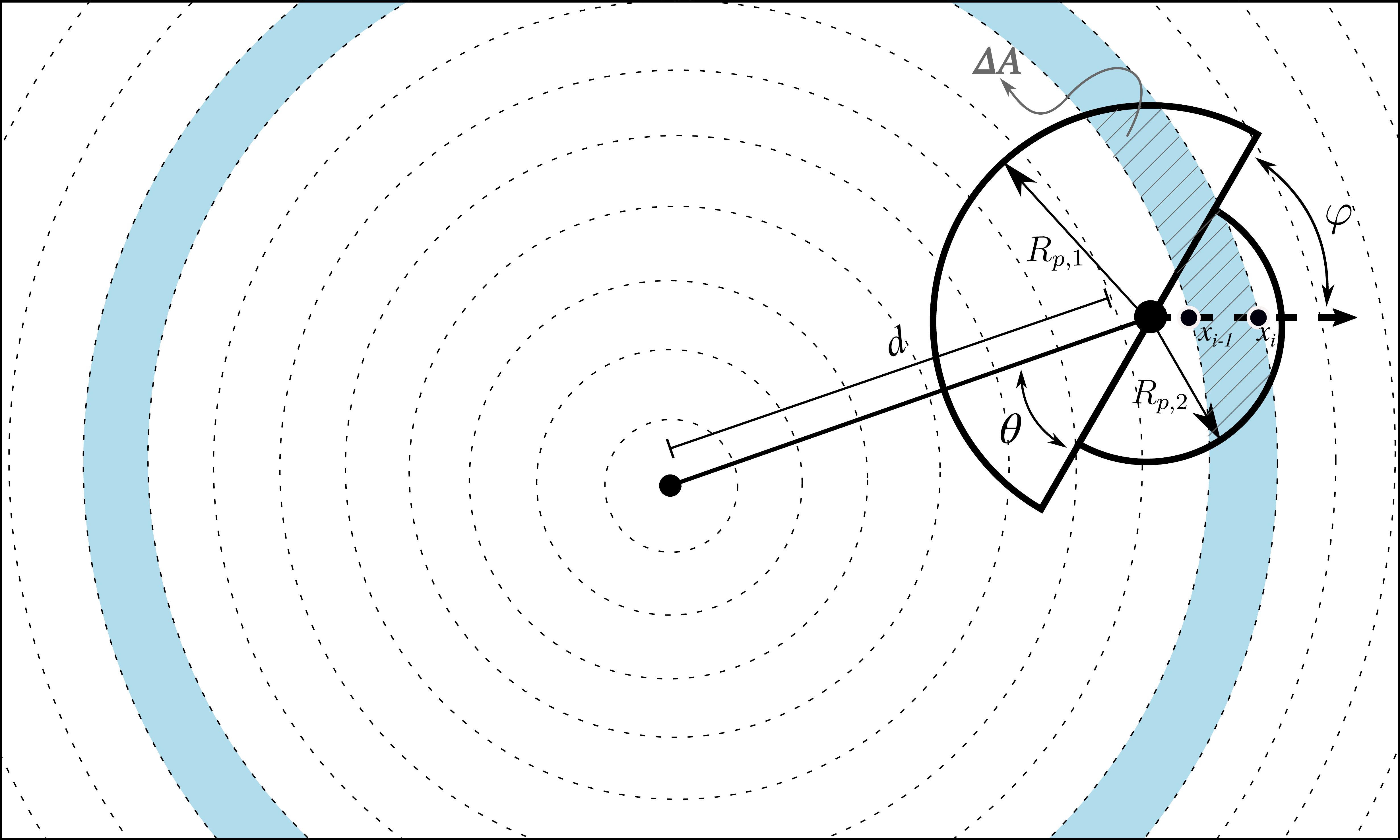}
    \caption{Diagram of the geometric configuration during transit of two stacked semi-circles (one of radius $R_{p,1}$; indicated by the arrow going up, and another of radius $R_{p,2}$; indicated by the arrow going down) that model the (possible) different limbs of an exoplanet transiting in front of a star. The area of the star has been divided into different sections of radius $x_i$ (dashed circles) --- between each subsequent section, the star is assumed to have a radially symmetric intensity profile (e.g., blue band between $x_{i-1}$ and $x_i$ above). In order to obtain the lightcurve such an object would produce, the challenge is to calculate the intersectional area between a given iso-intensity band and the stacked semi-circles, $\Delta A$ (blue band with dashed grey lines). Note the stacked semi-circles are inclined by an angle $\varphi$ with respect to the planetary orbital motion (illustrated by the dashed arrow moving to the right), which accounts for the possibility of having planetary spin-orbit misalignments {{($\varphi = \pi/2$ implies no spin-orbit misalignment)}. $\theta$ is the angle between the 
    base of the semi-circles and the line that joins the centers, $d$.}}
    \label{fig:geometry}
\end{figure*}

The basic problem we are trying to tackle is that of producing a transit lightcurve of two stacked semi-circles of (normalized, with respect to the 
stellar) radii $R_{p,1}$ and $R_{p,1}$ in 
front of their host star, where the semi-circles may 
be inclined with respect to the orbital motion by 
an angle $\varphi$. The geometrical configuration 
of the problem is depicted in Figure \ref{fig:geometry}. We follow \cite{batman} and assume 
a radially symmetric intensity profile $I(x)$, where 
$0<x<1$ is the normalized radial coordinate measured from the center of the star. With this, we can 
express the fraction of stellar light blocked by 
the object, $\delta$, as (see Figure \ref{fig:geometry})
\begin{eqnarray}
\label{eq:theproblem}
\delta = \sum_{i=1}^{N} I\left(x_m\right)\Delta A(x_m,R_{p,1},R_{p,2},\varphi,d),
\end{eqnarray}
where $x_m = (x_i + x_{i-1})/2$ is the middle point between $x_i$ and $x_{i-1}$, and $\Delta A(x_m,R_{p,1},R_{p,2},\varphi,d)$ (for which we 
shall refer to in what follows simply as $\Delta A$) is the 
inter-sectional area between the stacked semi-circles 
and the iso-intensity band depicted in Figure 
\ref{fig:geometry}, where $R_{p,1}$ and $R_{p,2}$ are the radii of the semi-circles, $\varphi$ is the rotation of the base of the semi-circles with respect to the orbital motion of the planet and $d$ is the distance between the center of the star and the semi-circles. Because the form of 
$I\left(x_m\right)$ is usually known/parametrized 
via so-called limb-darkening laws, the challenge 
of finding the lightcurve of this configuration of 
stacked, rotated semi-circles is to find $\Delta A$. The full derivation of this is presented in 
Appendix \ref{sec:deltaA}; we present an overview of our implementation and validation of our approach below.

\subsection{Implementation and model overview}
\label{sec:catwoman}

Our semi-analytic approach to the problem has been implemented in the \texttt{catwoman} 
library \citep{jones}, which is fully documented\footnote{\url{http://catwoman.readthedocs.io}} and available on 
Github\footnote{\url{https://github.com/KathrynJones1/catwoman}}. In practice, \texttt{catwoman}'s code-base 
is that of \texttt{batman} \cite{batman}, and as such the library inherits most of the high-level functionalities 
of this latter library. A \texttt{catwoman} lightcurve, thus, receives as inputs the time-of-transit center 
$t_0$, the period $P$ of the orbit, the scaled semi-major axis $a/R_*$, the inclination $i$ of the orbit 
with respect of the plane of the sky, the eccentricity $e$ and argument of periastron $\omega$ of the orbit, 
and a set of limb-darkening coefficients for any of the laws already available in 
\texttt{batman}. On top of these, \texttt{catwoman} takes as input the radii of each of the 
stacked semi-circles, $R_{p,1}$ and $R_{p,2}$, and the angle $\varphi$ between the axis that connects them and 
the vector that follows the direction of motion in the orbit (see Figure 1). 

The motivation behind allowing to define the angle $\varphi$ in the lightcurve generation comes from the 
possibility of being able to detect the sky-projected spin-orbit misalignment of the \textit{planet}, which is something 
the eclipse mapping technique for both lightcurves \citep{r7,w6} and radial-velocities \citep{n15} are able to do 
in principle. As will be shown in Section \ref{sec:detecting-varphi}, detecting the effect of asymmetric lightcurves 
due to morning/evening terminator structural and/or compositional inhomogeneities almost guarantees the possibility 
of putting constraints on this angle, and ignorance on its value 
does not have a great impact on the detectability of the effect. One important point to consider on this parameter is that this defines the 
\textit{instantaneous} angle between the axis that joins the semi-circles and the direction of the orbital 
motion (see Figure \ref{fig:geometry}; orbital motion indicated with a dashed-line arrow). Because orbits as projected in the plane of the sky are curved in general, this means the axis that joins 
the semi-circles \textit{rotates} when compared against a straight line projected in this plane. This effect has been 
implemented within \texttt{catwoman} as well (see Appendix \ref{sec:deltaA}); we validate this implementation 
against a numerical implementation in the next sub-section.

\subsection{Validation of the semi-analytical approach}
\label{sec:validation}
\begin{figure*}
\includegraphics[width=2.1\columnwidth]{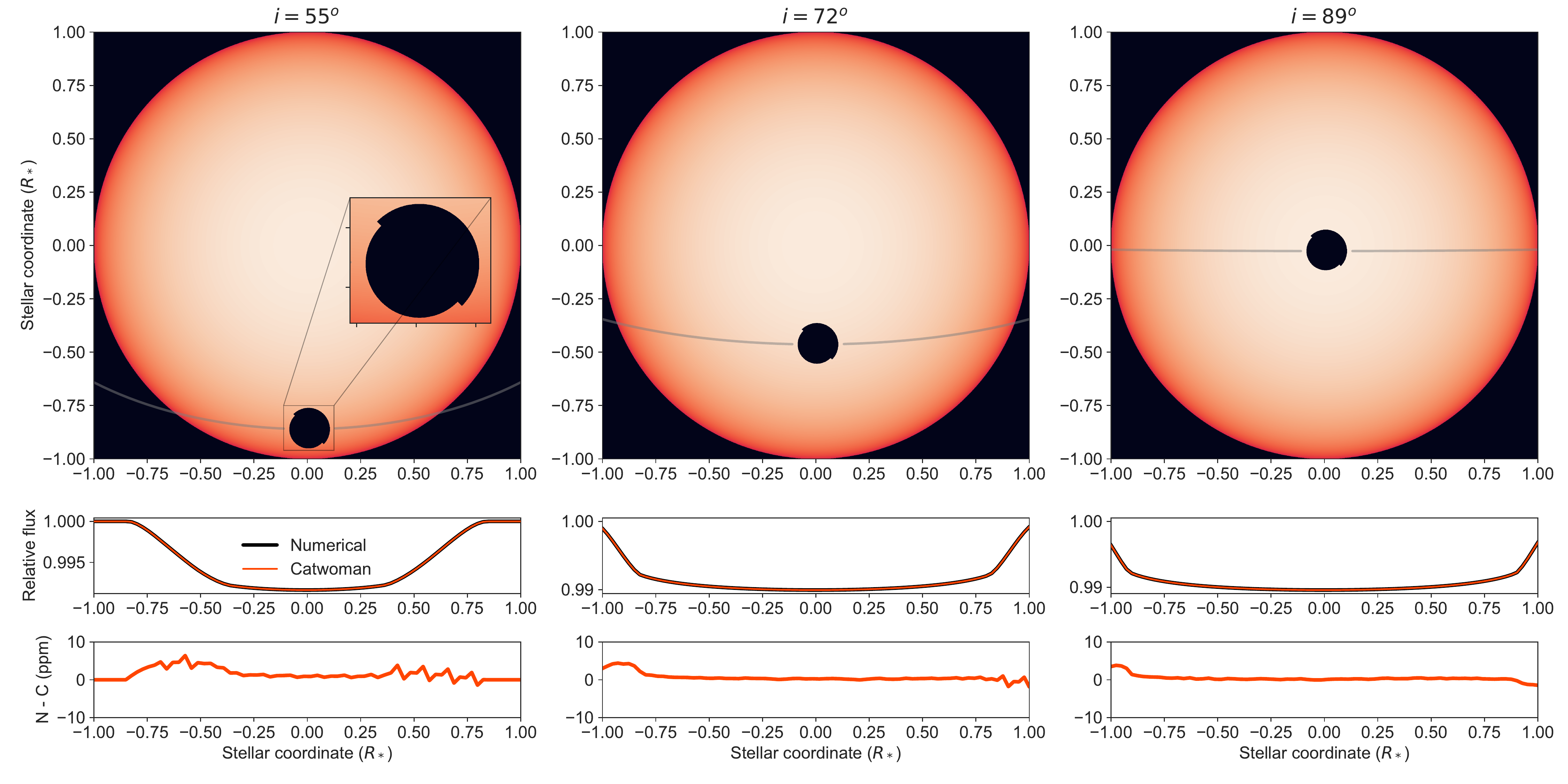}
    \caption{Comparison between a numerical implementation of the lightcurve generation of asymmetrical transits and the 
    semi-analytical formalism presented in our work, implemented in the \texttt{catwoman} library. Examples are 
    shown for orbital inclinations of 55$^o$ (left), 72$^o$ (middle) and 89$^o$ (right) --- all of them assume a period of 3 days, a quadratic limb-darkening law ($u_1 = 0.3$, $u_2=0.2$), $a/R_*=1.5$ 
    and zero eccentricity. The top images are snapshots 
    of our numerical model which include a limb-darkened star (orange) and a planet with asymmetric 
    terminator regions ($R_{p,1} = 0.1$, $R_{p,2} = 0.09$ and $\varphi = -45^{o}$) transiting in front of it; middle panels show the retrieved lightcurves from both methods, and the bottom 
    panels show the difference between the two. Most of the residuals observed in this latter panel are due to 
    errors on our numerical model scheme (see text); by construction, our \texttt{catwoman} models in these computations had a 1 ppm error limit.}
    \label{fig:numerical}
\end{figure*}

In order to validate the semi-analytical approach presented here and implemented in the \texttt{catwoman} 
library, we built a numerical model that is also able to generate asymmetric lightcurves due to terminator 
inhomogeneities but through a completely independent and straightforward (albeit ``brute-force") approach. {{While by construction \texttt{catwoman} is able to reach any desired 
precision level (as that is a parameter that can be modified and is tested for convergence 
before running lightcurve model evaluations), our objective with this alternative approach is 
to validate and illustrate that \texttt{catwoman} is indeed able to generate asymmetric 
lightcurves with accuracies of at least 10 ppm, which are the noise limits 
ultra-precise spectrophotometers like the upcoming \textit{JWST} will be able to reach \citep{greene:2016}}.} Our implementation of this numerical scheme is also available in Github\footnote{\url{https://github.com/nespinoza/numerical_catwoman}}, and is detailed below. 

Our approach of this numerical version of the lightcurve generation of asymmetrical transits is very similar 
to that of \cite{vParis}, and consists of simply discretizing the plane of the sky into $n_p \times  n_p$ ``pixels" 
centered around the target star. Pixels within the star are filled with values between 0 and 1 according to a 
given intensity profile $I(\mu)/I(1)$, while positions that include either the planet or the sky are filled 
with zeroes. The precision on the lightcurves generated by this scheme, thus, can be optimized by simply increasing 
$n_p$. In practice, this is implemented by populating a matrix of dimensions ($n_p,n_p$), on which we first 
fill all pixels within a distance of $n_p/2$ from the center of this matrix (i.e., $(n_p/2,n_p/2)$) with intensities given by the defined intensity profile (a quadratic law in the case of our 
implementation) --- all other pixels are filled with zeroes. With this, we sum all the pixel values to 
compute our out-of-transit flux. Our algorithm then, using as inputs the coordinates of the center of 
the planet with respect to a reference frame centered on the star $(X,Y)$ at each time-step 
and the input angle $\varphi$, computes the slope of the orbital motion $s = dY/dX$ by simple differences at each time-step $i$, i.e., $s_i=(Y_{i+1} - Y_i)/(X_{i+1}-X_i)$. This is then used to 
compute the instantaneous rotation of the axis that joins the stacked semi-circles with 
respect to the orthogonal system that defines the $(X,Y)$ positions as 
$\arctan(s_i) + \varphi$. This axis is then used to separate the areas covered by both semi-circles, pixels inside of which are set to zero. 

We use this simple numerical scheme to validate the semi-analytical framework developed in this work by 
computing a set of cases including a challenging one in which the planetary orbit is significantly curved. 
This latter case allows us, in turn, to verify that our method outlined in Appendix \ref{sec:deltaA} is 
correctly accounting for the rotation of the axis that joins the semi-circles with respect to the 
orthogonal system that defines the $(X,Y)$ positions of the planet. To generate this, we simulate an 
exoplanet with a period of 3 days, time-of-transit center $t_0=0$, scaled semi-major axis $a/R_*=1.5$ and zero eccentricity for three inclinations: $i=55$, $72$ 
and $89$ degrees. For the star, we define a quadratic limb-darkening profile with $u_1 = 0.3$ and $u_2=0.2$. 
As for the physical properties of the planet, we assume it to have asymmetric terminator regions with 
$R_{p,1} = 0.1$, $R_{p,2} = 0.09$ and $\varphi = -45^{o}$. Planetary positions $(X,Y)$ were obtained using 
\texttt{catwoman} (which uses the exact same method as \texttt{batman} to calculate them) for 100 equally 
spaced timestamps between -0.5 and 0.5 days. We performed numerical simulations with 
$n_p = 2500,\ 5000,\ 10000,\ 20000$ and $40000$ (i.e., doubling the number of pixels on each side of 
our matrix), and found that the maximum flux changes roughly halved as well between each of those runs. These changes reached 4 ppm between $n_p=20,000$ and $n_p=40,000$, 
which we consider as our maximum error on the fluxes of our numerical scheme when selecting this latter number for $n_p$. Simulations using both 
our numerical (with $n_p = 40,000$) and semi-analytical (through the \texttt{catwoman} library, with a maximum error set to 1 ppm\footnote{{{In \texttt{catwoman}, we inherit the maximum allowable truncation error for numerical integration 
from \texttt{batman}; see Section 3.4 in \cite{batman}}}.}) schemes are presented in Figure \ref{fig:numerical}. As can be observed, the differences between both are \textit{very} small; they reach peak differences of less than 7 ppm --- most of which 
are explained by the errors defined by our numerical scheme. 

For all practical purposes, these limits give 
us confidence that our semi-analytical framework works as expected for precisions which are better 
than current and near-future instruments such as \textit{JWST}, which is expected to reach about 10 ppm lightcurve precisions \citep{greene:2016}. We note that the speed increase of the \texttt{catwoman} 
library in comparison to the numerical implementation is huge: \texttt{catwoman} takes a couple hundreds of 
milliseconds to generate a lightcurve in a 2.9 GHz Intel Core i9 processor. The numerical implementation takes tens 
of seconds to generate the same model{{, although this latter one is a pure-Python code, whereas the 
\texttt{catwoman} library is a mixture of Python and C}}. In general, in experiments made with this processor, \texttt{catwoman} takes 
about twice the time \texttt{batman} takes to generate a lightcurve. This is consistent with the fact that the 
\texttt{catwoman} code-base is inherited from the \texttt{batman} one, and goes to show that the analytical part of 
\texttt{catwoman} is as fast as \texttt{batman}'s --- only that we perform it twice, one for each of the 
stacked semi-circles.

%%%%%%%%%%%%%%%%%%%%%%%%%%%%%%%%%%%%%%%%%%%%%%%%%%
%%%%%%%%%%%%%%%%%%%%%%%%%%%%%%%%%%%%%%%%%%%%%%%%%%
%%%%%%%%%%%%%%%%%%%%%%%%%%%%%%%%%%%%%%%%%%%%%%%%%%
%%%%%%%%%%%%%%%%%%%%%%%%%%%%%%%%%%%%%%%%%%%%%%%%%%
\section{Detectability of the effect}
\label{sec:detection}
Although the pioneering study of \cite{vParis} already tried to detect the effect of asymmetric 
transit lightcurves produced by non-uniform cloud cover on precise data of three 
exoplanets obtained by the \textit{Kepler} mission and \textit{HST}, a systematic 
study of the detectability of the effect has not been done either on real or simulated 
transit lightcurves. Such a study is very timely as the \textit{TESS} satellite 
\citep{ricker-tess} has just started its extended mission re-observing some of the most 
promising targets to detect this effect and as \textit{JWST} prepares for launch. These 
missions have a key advantage over \textit{Kepler}: they allow us to target objects with 
large scale-heights, for which this effect should be more prominent in the data even if 
they are observed over shorter time-scales. 

The question of the detectability of the effect in a given dataset is, however, a complex one. 
It is not only related to the precision of the lighcurves themselves in order to be able to 
detect the effect (which is evident will depend on the difference between the effective size 
of the terminator region on the leading and trailing limb of the exoplanet), but also to 
the correlation between the parameters that could impact on a transit lightcurve. It could be 
that the lightcurve indeed is asymmetric but a given transit parameter is able to correct for 
this if a symmetric model is used. Indeed, \cite{vParis} identified that the evidence for 
asymmetric lightcurves is heavily impacted by the knowledge of the ephemerides: a small shift 
in the time-of-transit center on a symmetric transit model could lead to an equally good fit 
to one with an asymmetric model, even for intrinsically asymmetric lightcurves. As such, in 
order to \textit{claim} the detection of this effect, one needs to perform proper model comparison. 
In this work, we choose to use bayesian model evidences to this end. In particular, we assume 
both the symmetric and asymmetric lightcurve models are equiprobable a-priori, which implies 
the difference between the log-evidence of an asymmetric lightcurve, $\ln Z_A$ to the one 
obtained from a symmetric one, $\ln Z_S$, $\Delta \ln Z = \ln Z_A - \ln Z_S$, is equal to 
\begin{eqnarray*}
\Delta \ln Z = \ln \frac{Z_{A}}{Z_{S}} = \ln \frac{\mathcal{P}(A | \textnormal{Data})}{\mathcal{P}(S | \textnormal{Data})},
\end{eqnarray*}
where $\mathcal{P}(A | \textnormal{Data})$ is the probability of the asymmetric model given 
the data and $\mathcal{P}(S | \textnormal{Data})$ is the probability of the symmetric model 
given the data.

In what follows, we simulate asymmetric lightcurves using the 
\texttt{catwoman} library with \textit{JWST}-like and \textit{TESS}-like cadences, 
in order to study how the detectability of the effect changes with our knowledge of different 
parameters of the model and the lightcurve precision using bayesian evidences 
as the metric for detectability. We decide not to generate simulations for \textit{HST}, as the gaps between 
orbits of the observatory imply a special, case-by-case analysis on the detectability of the effect --- we 
leave such a study for future work. For each of the cases described below we generate asymmetric 
lightcurves with a range of radius differences between the leading (``morning") and 
trailing (``evening") limbs. We parametrize this in our simulations in terms of the 
corresponding ``transit depth" each side of the planet implies. To this end, we fix 
$R_{p,1}/R_*$ to 0.1 in order to emulate a typical hot Jupiter planet-to-star radius ratio, 
and then define{{
\begin{equation}
\label{eq:me-def}
R_{p,2}/R_* = \sqrt{\left(R_{p,1}/R_*\right)^2 + 2\Delta \delta},
\end{equation}
where 
$\Delta \delta$ is the morning-to-evening transit depth difference. The factor of $2$ in front of this term stems 
from the fact that here we define $\Delta \delta$ as the transit depth difference \textit{between the transit depth 
imprinted in the lightcurve by each of the stacked semi-circles.} In other words, $\Delta \delta = \delta_2 - \delta_1$, where $\delta_i = (1/2)(R^2_{p,i}/R^2_*$). {{While there 
is no consensus in the literature as to how small or large morning-to-evening transit depth 
differences should be (e.g., \cite{kempton:2017} predict between 100-400 ppm differences for 
WASP-121b; \cite{powell:2019} predict values as large as 1000 ppm) we choose here to 
take a conservative upper limit on the effect of 500 ppm;}} in our simulations, 
thus, $\Delta \delta$ ranges from 5 to 500 ppm in 30 log-spaced bins}}. For each of those combinations, 
we simulate 
5 datasets of noisy transit lightcurves with noise levels $\sigma_w$ ranging from 10 to 
1000 ppm in 30 log-spaced bins as well. We calculate the average of the log-evidences 
for symmetric and asymmetric models fitted to that data in each $(\Delta \delta,\sigma_w)$ 
pair, which is then used to compute the difference between the log-evidences. In all of 
our simulations the period is set to 1 day, the semi-major axis to stellar radius ratio to 
$a/R_* = 10$, inclination to 90 degrees, and a circular orbit is assumed. We note this set of 
parameters define a worst-case scenario for the detection of the effect. The reason is that most 
of the information used to infer the limb asymmetries comes from ingress and egress, as has already 
been shown by previous works \citep[see, e.g.,][]{vParis, kempton:2017, powell:2019}. The ingress/egress 
duration in a circular orbit is given by
\begin{eqnarray*}
\tau = \left(\frac{P}{\pi}\right)\left(\frac{1}{\sqrt{1-b^2}}\right)\left( \frac{R_p}{R_*}\right)\left(\frac{R_*}{a}\right).
\end{eqnarray*}
In the case of these simulations, this gives an ingress/egress duration of only $\tau = 4.6$ minutes. As a 
comparison, the archetypal hot Jupiter HD 209458b \citep{c:2000,h:2000} has $\tau = 25.7$ minutes. Our simulations in 
this Section, thus, can be seen as \textit{lower limits {{and/or very 
conservative estimates}}} on the detectability of the effect. We explore the variation 
of the precision on the limb asymmetries with ingress/egress duration along with a case-study of a real hot Jupiter 
in Section \ref{sec:discussion}.

To perform the fits to our simulated data, we implemented \texttt{catwoman} \citep{jones} in 
the \texttt{juliet} \citep{juliet} package, which already implements 
\texttt{batman} \citep{batman} for symmetric lightcurve models, and allows us to compute bayesian evidences 
for our model comparison using MultiNest \citep{MultiNest} via the PyMultiNest wrapper \citep{PyMultiNest}. {{Table \ref{tab:priors1} lists the prior distributions 
used in our experiments --- we explain and detail each of those below.}}

\begin{deluxetable}{lrl}[b!]
\tablecaption{Priors and parameters used for our experiments in Section 3 for the \texttt{catwoman} model fits. The \texttt{batman} fits used similar priors, with the caveat that this model 
assigns a uniform radius for the entire planet $R_p/R_*$ for which we use the same prior as for $R_{p,1}/R_*$ below, and doesn't fit for $\varphi$. $U(a,b)$ below stands for a uniform distribution between $a$ and $b$.\label{tab:priors1}}
\tablecolumns{3}
\tablenum{1}
\tablewidth{1pt}
\tablehead{
\colhead{Parameter} &
\colhead{Prior} & 
\colhead{Comment}\\
}
\startdata
P (days) & --- & Fixed to 1 \\
T$_0$ (days)&  $U(-0.1,0.1)$ & Fixed to 0 when\\
 & & assumed known.\\
$R_{p,1}/R_*$ & $U(0, 1)$ \\
$R_{p,2}/R_*$ & $U(0, 1)$ \\
$\varphi$ (deg) & $U(-90, 90)$ & Only used in Section \ref{sec:detecting-varphi}. \\
 & &  Fixed to 90 otherwise.\\
$q_1$\tablenotemark{a} & $U(0, 1)$ & Fixed to 0.25 when\\
 & &  assumed known.\\
$q_2$\tablenotemark{a} & $U(0, 1)$ & Fixed to 0.3 when\\
 & & assumed known.\\
$e$ & --- & Fixed to 0 \\
$\omega$ & --- & Fixed to 90 \\
$a/R_*$ & --- & Fixed to 10 \\
$b$ & --- & Fixed to 0 \\
%r1\tablenotemark{a} & $U(0,1)$ & \rone \\
%r2\tablenotemark{a} & $U(0,1)$ & \rtwo \\
\enddata
\tablenotetext{a}{These parameters correspond to the parametrization presented in \citet{kipping:2013} for sampling physically plausible combinations of the quadratic 
limb-darkening coefficients. }
%\tablenotetext{b}{Time-averaged equilibrium temperature computed according to equation~16 of \citet{mendez:2017}}
\end{deluxetable}

\subsection{Detecting asymmetric lightcurves with \textit{JWST}}
\label{sec:jwst-like}

In order to perform the simulations for \textit{JWST}-like observations, we needed to 
calculate a typical cadence for observations to be taken by the observatory for 
time-series exposures. In this work, we focus on observations aiming to constrain the effect using 
NIRISS/SOSS, as this instrument allows us to obtain spectra all the way down to $0.6\ \mu m$ through 
the combination of data from Order 1 ($1-3\ \mu m$) and 2 ($0.6-1\ \mu m$). Given that the largest limb 
asymmetries seem to be in the transition between optical and NIR wavelengths \citep[see, e.g., ][]{kempton:2017, powell:2019}, we believe this will usually be the instrument of choice to characterize the effect for bright targets (with the alternative 
being, of course, NIRSpec/PRISM for fainter targets). 
Considering the reset time for this instrument is relatively short (couple of seconds), it sufficed for 
our work to know the typical integration time of a \textit{JWST} observation with SOSS. For a solar-type, $V=11$ star, 
according to the \textit{JWST} Exposure Time Calculator\footnote{\url{https://jwst.etc.stsci.edu/}} 
\citep[ETC; ][]{jwst-etc}, ``saturation" is attained at about 20-60 groups per integration for NIRISS/SOSS, which implies maximum integration times 
between 40-80 seconds per datapoint in the time-series. We arbitrarily decided to use 20 
seconds for the cadence of our simulations in order to simulate observations trying to 
target half-saturation values, which has been a typical strategy for \textit{HST} observations\footnote{But see \url{https://jwst-docs.stsci.edu/methods-and-roadmaps/jwst-time-series-observations/tso-saturation}.}.

\begin{figure*}
\includegraphics[width=2.1\columnwidth]{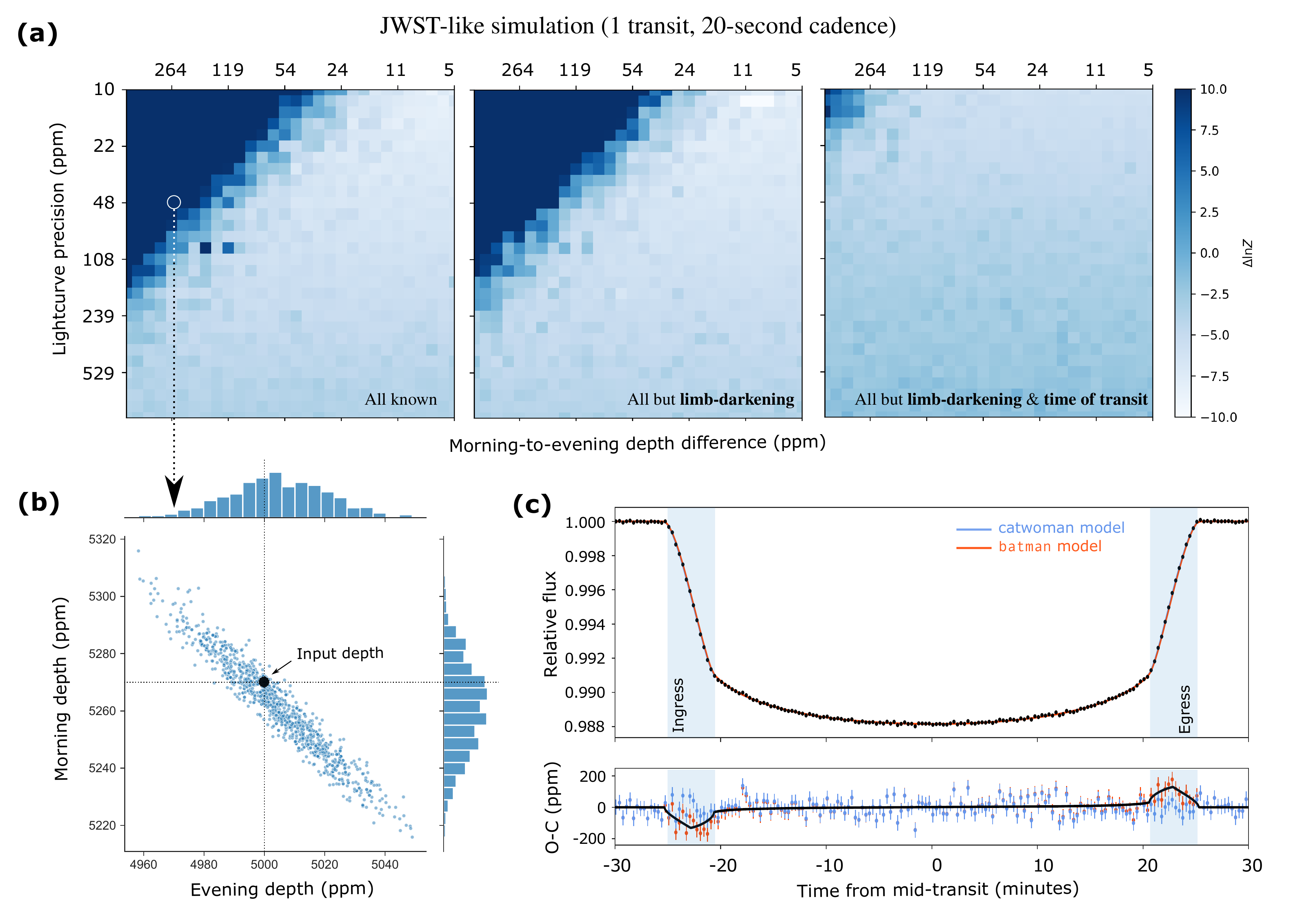}
    \caption{{{\textbf{Injection and recovery simulations of asymmetric lightcurves due to differences in the morning and evening terminator for \textit{JWST}-like cadence (1 transit, 20-second cadence, 1-day period, 4.6-minute ingress/egress duration)}. \textbf{(a)} Detection 
    map in the morning-to-evening depth difference versus lightcurve precision (both in ppm) map; 
    darker regions show the detectability region of the effect}} when all the parameters other than the planetary radius is known (left), other than the planetary radius and limb-darkening (center) and other than the planetary radius, limb-darkening and 
    time-of-transit center (right). Colors indicate the difference between the log evidences{{, $\Delta \ln Z$,}} of asymmetric {{(\texttt{catwoman})}} models and symmetric {{(\texttt{batman})}} models. Note how if the time-of-transit center is unknown{,  detecting the effect gets very challenging. \textbf{(b)} Posterior samples of the morning and evening limb depths on an example lightcurve with 50 ppm precision and a morning-to-evening depth difference of 270 ppm; note the high correlation (but good 
    recovery) of the limb depths; true input value is marked with dashed lines. \textbf{(c)} Simulated 
    transit lightcurve corresponding to the posterior shown in (b); models (\texttt{batman} in 
    orangered; \texttt{catwoman} in blue) are indistinguishable in the top panel; the residual panel, 
    however, clearly shows the differences: the residuals using the \texttt{batman} model (orangered) 
    show bumps at ingress and egress; the \texttt{catwoman} model residuals (blue) correctly 
    model those bumps. The black line in the bottom panel shows the difference between the best-fit \texttt{batman} and \texttt{catwoman} models.}}
    \label{fig:jwst-like}
\end{figure*}

In this case we tried three different simulations, in order to illustrate the impact of 
different assumptions in this one-transit, 20-second cadence case: (a) one in which 
everything but the radii of both sides of the exoplanet $R_{p,1}$ and $R_{p,2}$ are known, 
(b) one in which everything but the radii and the the limb-darkening coefficients of the 
star are known, and finally (c) one in which everything but the radii, the 
limb-darkening coefficients of the star and the time-of-transit center are known. The 
limb-darkening law that was assumed to generate and fit the lightcurves was a 
quadratic law with $(u_1,u_2) = (0.3,0.2)$, which are representative values for solar-type 
stars. In our \texttt{juliet} fits, we assumed the uninformative priors for two-parameter 
limb-darkening laws proposed by \citep{kipping:2013} for the cases in which limb-darkening 
was assumed to be unknown; for the time-of-transit center, a uniform prior with a width of 
{{4.8}} hours around the real predicted time of transit center was imposed. The results for 
these simulations are presented in Figure \ref{fig:jwst-like}.

As can be observed from the simulations, the asymmetric transit lightcurves should be 
detectable (i.e., $\Delta \ln Z \gtrsim 2$) for morning-to-evening depth differences 
above around {{$25$}} ppm for a wide range of precisions at least in white-light (i.e., adding all the flux 
over the entire wavelength range of a given instrument), 
where \textit{JWST} observations should achieve tens of ppm precisions per point in the transit 
lightcurves, and as long as the ephemerides are well known and constrained. If they are 
not, however, as can be observed in the rightmost panel of Figure \ref{fig:jwst-like}{{a}}, 
the actual detection of the effect becomes extremely challenging because, as it has been 
already noted by \cite{vParis}, changes in the time-of-transit center in a symmetric model 
can account for the asymmetry in the lightcurve. The changes in this timing are very small --- only a couple of seconds of shifts in the time-of-transit center suffice to mimic 
the asymmetry in the transit lightcurves (see Section \ref{sec:timing} for details). This implies that to detect this effect 
in white-light, very precise timings are needed in order to claim a detection.

It is important to note that although from the above results the detection of the 
effect \textit{directly} in the white-light lightcurves even with \textit{JWST}-like precisions seems {{relatively}} challenging 
to do with only one transit in the absence of precise timing constraints, the observatory has the 
advantage that it can perform spectro-photometry and, thus, the effect can be detected 
through the wavelength dependence of the radii at each side of the terminator region, as 
has already been highlighted by \cite{powell:2019} --- see also Section 
\ref{sec:discussion}. In particular, NIRISS/SOSS can produce extremely precise (tens of 
ppm) white-light transit lightcurves in Orders 1 and 2, which can be used to claim a detection 
of the effect using these white-light transit lightcurves alone {{at much higher 
significance levels (and thus be sensitive to much lower morning-to-evening depth 
differences) than the ones shown here. Lightcurves like these, in addition, should provide 
very precise morning and evening depths. Figure \ref{fig:jwst-like}b and 
\ref{fig:jwst-like}c show an example lightcurve fit on a 50 ppm-precision lightcurve, where the 
injected morning-to-evening depth difference was of 270 ppm. As can be seen in Figure \ref{fig:jwst-like}b, both evening and morning depths are highly correlated, but nonetheless 
provide precise constraints in this case on each of about 16 ppm, giving in this case a retrieved 
morning-to-evening depth difference of $\Delta \delta = 257 \pm 32$ ppm --- fully consistent with the input 
value\footnote{Note the precisions of each limb do not add in quadrature to the constraint on 
the limb-difference. This is expected, again, due to the correlation between each of the 
limb depths.}. Most of the information to constrain those depths come from ingress and egress, as is 
evident in the residuals (blue for \texttt{catwoman}, red for \texttt{batman}) of Figure \ref{fig:jwst-like}c. We provide a deeper understanding on the relation between lightcurve 
precision and morning and evening depth precisions in Section \ref{sec:discussion}.}}

\subsection{Detecting asymmetric lightcurves with \textit{TESS}}

\begin{figure}
\includegraphics[width=1.05\columnwidth]{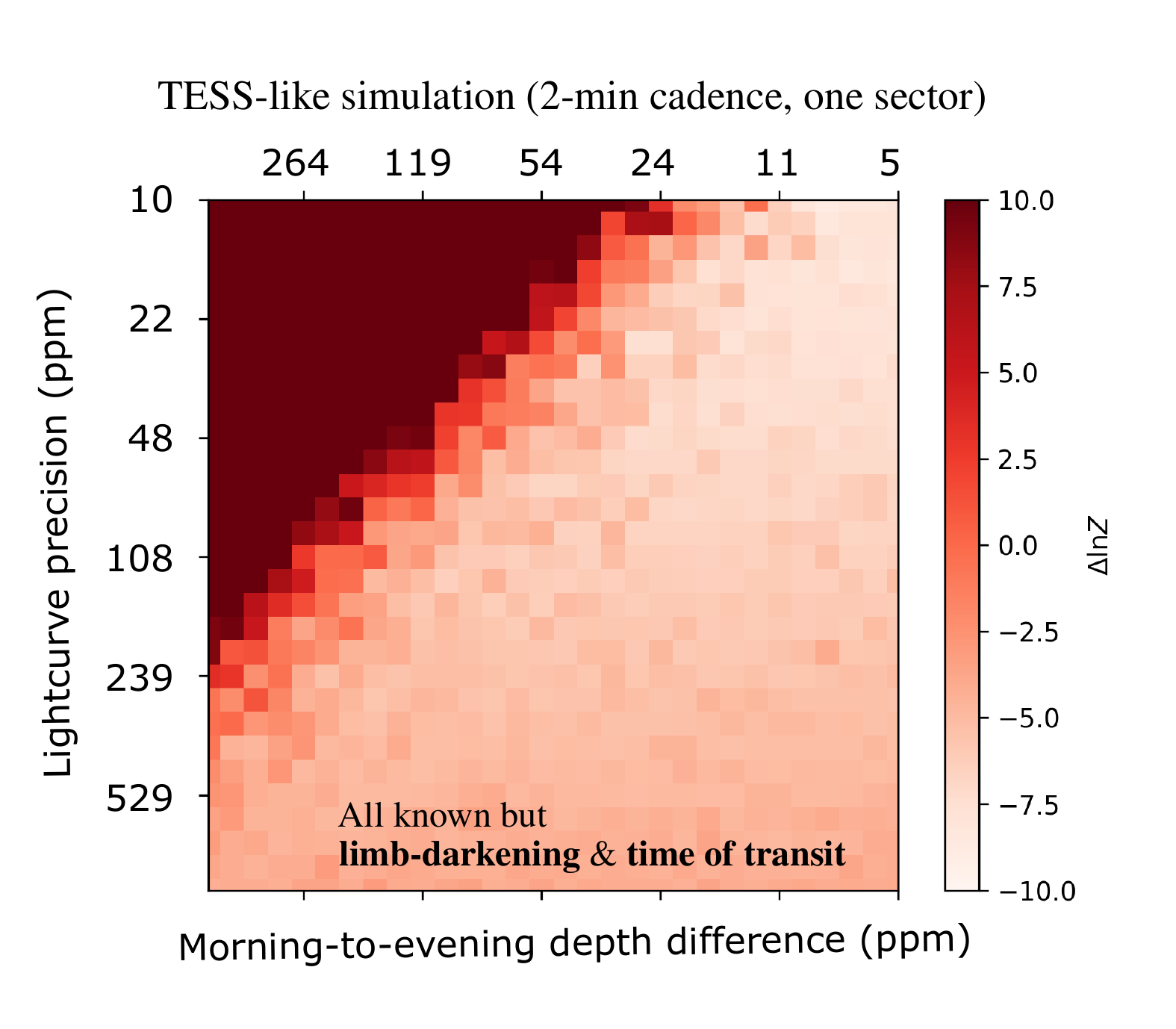}
    \caption{Simulations of asymmetric lightcurves due to differences in the morning and evening terminator transmission spectra for \textit{TESS}-like cadence (2-minute cadence, one sector, 1-day period, 4.6-minute ingress/egress duration) when all the parameters other than the planetary radius, limb-darkening and 
    time-of-transit center are known. Colors indicate the difference between the log evidences of asymmetric models and symmetric models (positive meaning odds ratios in favor of asymmetric lightcurve models).}
    \label{fig:tess-like}
\end{figure}

Although the \textit{TESS} mission has a significantly smaller aperture than \textit{JWST}, the 
cadence and types of observations the mission does are excellent for the detection of 
asymmetries in transit lightcurves. The mission not only attains an exquisite precision, 
but it is also able to observe several transits of the same exoplanet, {{mitigating}} the 
problem we observed with only one transit in \textit{JWST}-like observations like the 
ones simulated in the previous subsection. We performed the same simulations 
that we did for \textit{JWST} but with a \textit{TESS}-like cadence of 2-minutes, where 
we only consider observations on a 27-day period (i.e., one \textit{TESS} sector). 
Interestingly, the three cases that we tried in the previous sub-section (all known, all but 
limb-darkening known and all but limb-darkening and the time-of-transit center known) all 
resulted in practically identical results --- we show the one corresponding to the case in 
which all the parameters are assumed to be known but the radius, the limb-darkening 
coefficients and the time-of-transit center in Figure \ref{fig:tess-like}.

As can be observed, the results are very similar to the ones of \textit{JWST}. This is a combination 
of the fact that there is about a 27-fold increase in the number of transits, which helps 
with the 6-fold increase on the cadence of the observations as compared to the \textit{JWST} ones. The 
fact that there are more transits, in addition, helps with the problem \textit{JWST} will face 
related to the ephemerides where in our analysis, of course, there is an implicit assumption 
regarding no possible deviations from strict periodicity in the transit times. {{We 
reiterate, however, that our simple simulations in Section \ref{sec:jwst-like} did not consider the huge advantage that 
\textit{JWST} has over \textit{TESS} regarding the ability to measure wavelength-dependent 
transits, which should in turn break the degeneracy with the ephemerides as already suggested 
by \cite{powell:2019}. We delve deeper into the benefits of wavelength-dependent 
\textit{JWST} transit observations for detecting the limbs of exoplanets 
in Section \ref{sec:discussion}.}}

Although few targets attain the precisions at which one might statistically distinguish 
between an asymmetric and a symmetric model directly from the transit lightcurves with \textit{TESS} in one sector, 
the fact that many targets are observed by more than one sector makes this effect within 
reach of what \textit{TESS} is currently able to detect. Targets in the \textit{JWST} Continuous Viewing Zone (CVZ) are particularly 
appealing to try to detect this effect. {{For example, WASP-62b ($V=10.3$), for which the median per point precision was 
880 ppm during the prime mission in 2-minute cadence, has been observed to date in more than 20 sectors, providing a combined per-point 
precision per transit of about at least $880/\sqrt{20} \approx 200$ ppm --- a promising precision level to constrain the effect of 
asymmetric limbs if we assume a simulation like the one in Figure \ref{fig:tess-like} also applies to an exoplanet like WASP-62b. {{The recent \textit{HST} study by \cite{alam} makes this target also particularly appealing to detect this effect, as the atmospheric retrievals performed on its spectrum 
point it to have a colder temperature ($\approx$ 800 K) than what is expected from its equilibrium temperature ($\approx$ 1400 K). This is one of the key hints that there might indeed be differences between the limbs of this exoplanet, as suggested by the work of \cite{macdonald:2020}.}}}} 

\subsection{Detectability assuming $\varphi$ is not known}
\label{sec:detecting-varphi}

\begin{figure*}
\includegraphics[width=2.05\columnwidth]{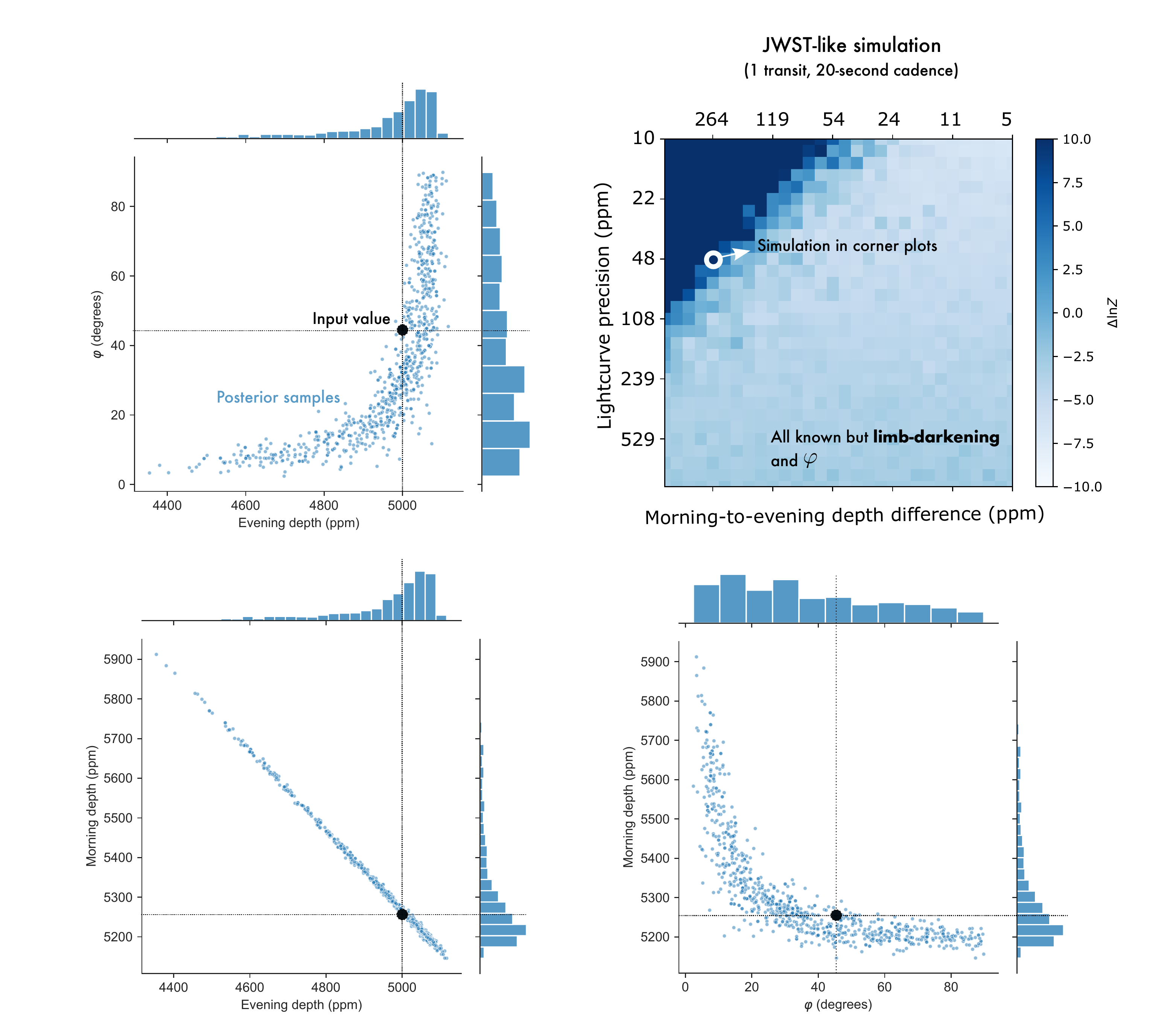}
    \caption{{{\textbf{Simulations allowing for an unknown projected planetary spin angle, $\varphi$}. The simulations presented above are similar to those shown on the center panel of Figure \ref{fig:jwst-like}a, but this time also allowing for $\varphi$ to be a free parameter. In addition to the detectability map, shown in the top right corner, we present 
    a portion of the posterior distribution (which has two modes, allowing for both positive 
    and negative values of $\varphi$; see text) of the limb depths and angle $\varphi$ as corner plots for the 
    very same limb difference (270 ppm) and lightcurve precision (50 ppm) as that shown in Figure \ref{fig:jwst-like} (indicated here as a white circle in the top right detectability plot. Black dot on the corner plots indicates the true input value.)}}}
    \label{fig:phi-not}
\end{figure*}
As a final test on the detectability of the effect, we explore whether our ignorance on the angle 
$\varphi$ can impact {{it}}; we take our \textit{JWST}-like simulation 
as a proxy for studying this, given the similarity in the shape of the detectability maps presented 
between \textit{JWST} and \textit{TESS} in Figures \ref{fig:jwst-like} and \ref{fig:tess-like}. To explore this, we 
use a transit lightcurve whose parameters are defined by the same ones as in the previous experiments, 
but in this case we set $\varphi = 45$ degrees as the ground-truth, and set a uniform prior between -90 and 90 
degrees for the parameter in our fits. Our results for a \textit{JWST}-like simulation (using the same cadence as in 
Section \ref{sec:jwst-like}) are shown in Figure \ref{fig:phi-not} {{where in 
addition to the detectability map, we also show a portion of the posterior distribution of 
a simulation with the same properties as the one shown in Figure \ref{fig:jwst-like} for 
the case in which $\varphi$ was fixed --- i.e., a lightcurve precision of 50 ppm, and a 
morning-to-evening depth difference of 270 ppm}}.

As can be seen, the detectability region (i.e., the medium blue and dark-blue region) of the 
plot has shifted by a small amount with respect to the one presented in Figure 
\ref{fig:jwst-like} implying that a slightly better lightcurve precision is needed in order to detect the effect if the 
angle $\varphi$ is not known a-priori. {{In addition, given our large prior on 
$\varphi$, we actually detect two posterior modes (of which we show only one in Figure \ref{fig:phi-not}): one allowing for positive and one allowing for negative values of 
$\varphi$, each swapping the posterior distributions between the evening and morning limb. 
This was expected by construction: a model with $\varphi$ and a given value of $R_{p,1}= a$ 
and $R_{p,2}=b$ is the same as a model with $-\varphi$ with $R_{p,1}= b$ 
and $R_{p,2}=a$. If we take the mode with positive values for $\varphi$, we note that in this case the evening and morning depths themselves are much more uncertain. This gives rise to a retrieved morning-to-evening 
depth difference of $250^{+374}_{-113}$ ppm --- again consistent, but much less precise than 
the constraint assuming a known value of $\varphi$ presented in Section \ref{sec:jwst-like}. 
The constraint on $\varphi$ itself is also not very precise; for this particular simulation, 
we obtain $\varphi = 32^{+33}_{-20}$ degrees; fully consistent with the input value, but not 
very constraining to understand the underlying, true projected planetary spin-angle. 

Before moving into the next Section, it is important to reiterate that the precisions and detectability limits shown here were obtained for a very conservative ---worst-case scenario--- system with a very small ingress/egress 
duration.}} As will be shown in Section \ref{sec:precisions}, the odds of detecting the 
effect on systems which have better prospects for it (i.e., systems with longer ingress/egress durations) 
are in reality much higher. The lower limits we set here, thus, seem promising for the detection of the effect 
with current and near-future instrumentation.

\section{Discussion}
\label{sec:discussion}

In previous sections, we have presented both the details of our semi-analytic framework 
for generating asymmetric transit lightcurves due to morning/evening terminator 
heterogeneities --- including its validation against simpler (but more computationally 
expensive) models --- and a study of the detectability of the effect with current 
missions like \textit{TESS} and future observatories like \textit{JWST}. Although our 
results are encouraging for the detection of the effect, there are many aspects to 
pay attention to when performing lightcurve analyses and/or when planning 
observations to detect the effect, including complementary methodologies, which 
we discuss below.

\subsection{Asymmetric terminator depths precision}
\label{sec:precisions}

While in Section \ref{sec:detection} we presented lower limits on the statistical 
detection of the effect on transit lightcurves based on bayesian evidences, 
an important aspect of interpreting transit lightcurve fits with the semi-analytic 
model presented in this work will involve constraining the actual measured transit 
depths of each side of the planet. This will be useful not only to extract transit 
spectra of the different limbs when using wavelength-dependent lightcurves such as 
the ones to be obtained by \textit{JWST}, but also to compute the maximum possible 
transit depth differences allowed by the data when analyzing broadband data such as 
the one from missions like \textit{TESS}. 

An important detail to consider when extracting transit depths from asymmetric limbs 
is the fact that the observable quantities that are directly constrained by the data 
are the areas of each of the semi-circles through the transit depths each of them 
produce. In symmetric models, where the limbs are assumed to be equal, the transit depth 
is simply $\delta = (R_p^2/R_*^2)$ --- the projected area of the planet 
over the projected area of the star. In the asymmetric case, however, the projected area 
of the \textit{semi-circles} are the quantities of interest, --- the transit depth 
of each limb being given by $\delta_i = (1/2)(R_{p,i}^2/R_*^2)$. This is what effectively 
defines the transit spectrum of each limb and is, in turn, what should be used to compare 
against theoretical transmission spectroscopy models. 

\begin{figure}
\includegraphics[width=1.08\columnwidth]{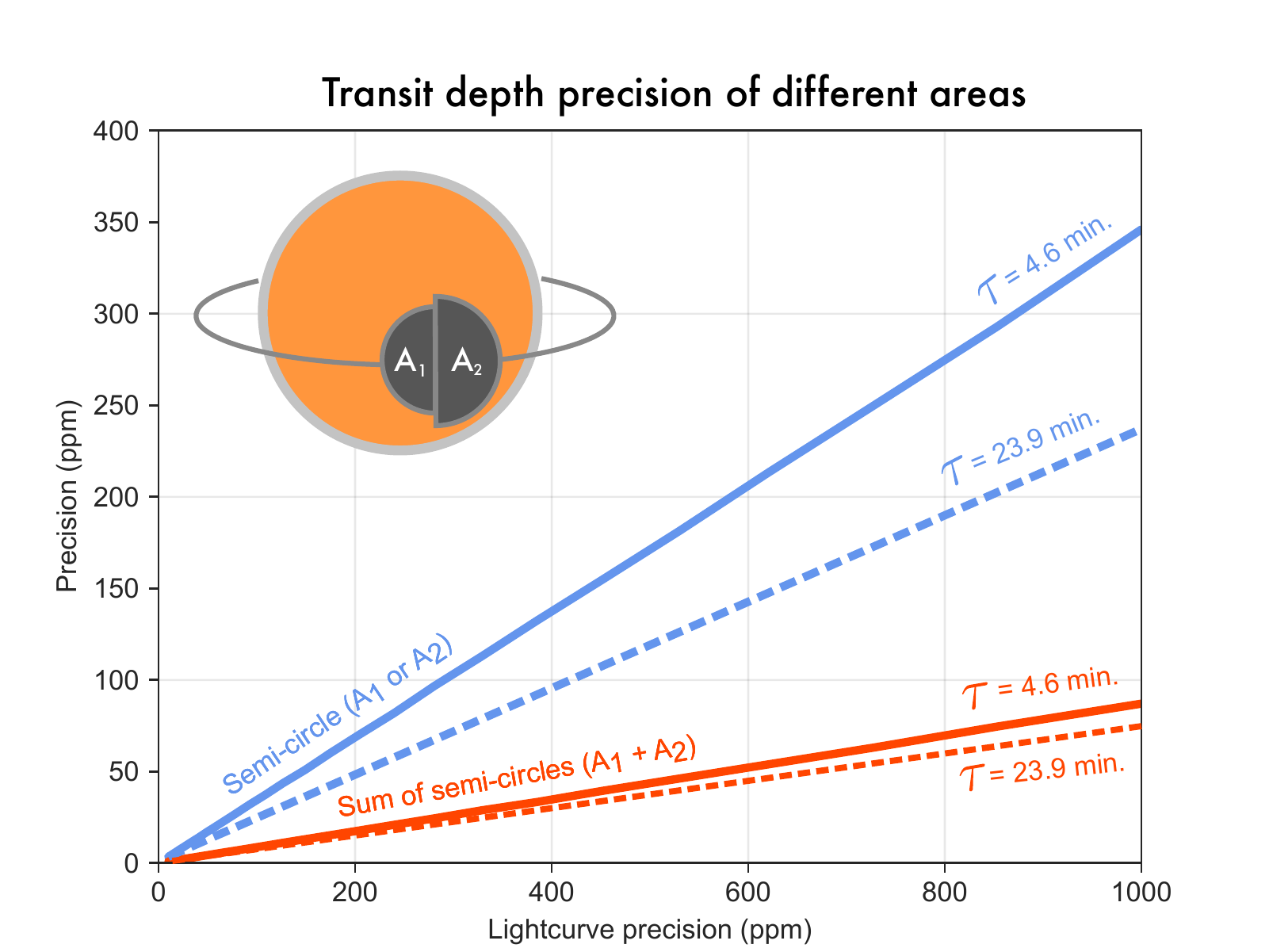}
    \caption{Transit depth precision on the semi-circles (transit depth defined as $\delta_i = (1/2)(R_{p,i}/R_*)^2$, with 
    $R_{p,i}$ being the area of semi-circle $i$; blue lines) and their sum (whose transit depth is $\delta_1+\delta_2$; red 
    lines) as a function of lightcurve precision. Precisions for systems with short 
    (4.6 min., solid lines) and long (23.9 min., typical for hot Jupiters) transit ingress/egress durations, $\tau$, are presented. \textbf{Main point:} Transit depth precisions of the limb of exoplanets are much less precise 
    and much more dependant on transit ingress/egress duration than the transit depth produced by the entire area 
    of the planet (i.e., the classically defined ``transit depth").}
    \label{fig:precision}
\end{figure}

Figure \ref{fig:precision} shows how the precision of the transit depths of each limb depend 
on the lightcurve precision, as well as the precision of the entire area of the planet, defined by the depth 
$\delta_1 + \delta_2$, for the case of the exoplanet simulated in Section \ref{sec:detection} with an 
ingress/egress duration of $\tau = 4.6$ minutes (solid lines). As can be seen, 
the precision on the transit depth of the entire planet is always much smaller 
than the corresponding for both semi-circles, but the relationship between the two is not 
simple, as the transit depths of the semi-circles are highly correlated with each other. 
Indeed, the transit depth of the entire planet is constrained by \textit{the entire 
lightcurve}, whereas the transit depths of each side of the planet (sampled by 
the semi-circles in our model) are mostly constrained by ingress and egress. This implies, 
in turn, that this latter precision would of course increase on systems with 
larger ingress/egress durations, which in many cases might be the optimal ones to target 
in order to maximize the chances to unveil this effect. 

The simulations presented in Section \ref{sec:detection} for $\tau = 4.6$ min. ingress/egress 
durations were performed to show lower-limits on the detection of the effect, and even in 
those cases the odds were very favorable given current (e.g., \textit{TESS}) and future 
(e.g. \textit{JWST}) lightcurve precisions and cadences. Hot Jupiters typically have 
longer ingress/egress durations, and some of the already characterized ones by missions like, e.g., 
\textit{HST}, show good prospects for the detection of the effect as well. As an example, 
we repeat the \textit{JWST} simulations in Section \ref{sec:detection} for HAT-P-41b, {{which we 
select as is one of the most thoroughly characterized ultra-hot Jupiters in 
transmission --- all the way from near-UV,}} optical and {{up to}} near-infrared wavelengths \citep{w2020, l2020, s2020}. We tune the physical and orbital parameters of the system 
to the ones used in \cite{w2020}, which imply a 23.9 minutes ingress/egress duration. The cadence (53 seconds) 
and number of datapoints (500) for our simulations are set to the ones optimized by PandExo\footnote{\url{https://exoctk.stsci.edu/pandexo/}} \citep{pandexo} using NIRISS/SOSS as the 
instrument of choice such that SUBSTRIP256 does not saturate (which would be the setup of choice in order to 
obtain simultaneous spectroscopy in the near-infrared and the optical through Orders 1 and 2 for this 
target\footnote{We note HAT-P-41b saturates below about 2 microns with NIRSpec/PRISM, which is the reason why 
we don't discuss this instrument in the context of this exoplanet.}). For consistency, we set the limb-darkening 
coefficients to the average ones on Order 1 of NIRISS/SOSS (but we note these do not impact on the overall precision and 
detectability, as was already shown in Section \ref{sec:detection}). The resulting precisions of this experiment are presented in Figure \ref{fig:precision} as dashed lines. As can be observed, the precision change on the 
transit depths of each of the limbs is significant, and ranges from a 60\% to 70\% improvement in it. The 
precision change in the transit depth of the entire planet, however, is much smaller (and driven mainly by the 
difference in the absolute transit depths and transit durations), which acts as a baseline in 
showing quantitatively how the prospects of detecting asymmetric limb-differences are very sensitive to the 
ingress/egress duration. 

One might argue that in Figure \ref{fig:precision}, one could observe two transits of the exoplanet with 
$4.6$ min. ingress/egress duration in order to match the signal-to-noise of the exoplanet with the 23.9 min. 
ingress/egress duration. Although this would be true if the observatory only targeted the transit event, in 
practice there are observational overheads (like, e.g., pre-post transit baselines, and overall observatory 
overheads beyond clock time on-target) that have to be included in that reasoning. For instance, the time 
recommended in the \textit{JWST} documentation\footnote{\url{https://jwst-docs.stsci.edu/}} one should spend 
in a target during a transit is given by the dwell equation, which reads
\begin{equation}
\label{tdwell}
T_{\textnormal{dwell}} = 0.75 + \textnormal{MAX}(2, T_{14}) + T_{14} + 1\ \textnormal{hr},
\end{equation}
where $T_{14}$ is the transit duration in hours. The exoplanet with an $4.6$ min ingress/egress duration has a 
1-hour total transit duration, which gives $T_{\textnormal{dwell}} = 4.75$. The exoplanet with a 23.9 min ingress/egress 
duration has a $3.6$-hour transit duration, $T_{\textnormal{dwell}} = 8.95$. Two transits of the 1-hour transit 
duration target would imply a requested time of $9.5$ hours, which is at least half an hour more expensive than the 
$3.6$-hour transit duration target --- all this without considering extra observatory overheads. The conclusion, thus, 
is that the efficiency of the time and targets to be requested to detect the effect have to be studied on a 
case-by-case basis.

{{In addition to the above, it is also important to note that the posterior distributions 
of the limbs show typically non-negligible correlations (see, e.g., Figures \ref{fig:jwst-like} and \ref{fig:phi-not}). This is important to consider because this covariance carries extra information that 
could become important when performing inference on the limbs (through, e.g., atmospheric retrievals). In our experiments, we observed the shape 
of the posterior distribution (measured through the correlation --- i.e., the covariance divided by the 
standard deviations of the marginal posterior distribution of each limb) remained fairly constant across the 
parameter space covered in our work; it simply shrinks as better precisions are achieved. We also found the 
posterior distribution of the limbs is very well approximated in the case of a known angle $\phi$ by a 
two-dimensional gaussian distribution, which is characterized not only by the mean and standard deviation of 
the marginal posterior distributions, but also by the covariance between the limbs. We provide a practical 
example of how to use this information in a retrieval framework in the next sub-section.}} 

\subsection{The importance of constraining limb spectra}
\label{sec:retrievals}

%, using the simulations described in 
%the previous sub-section. We use the PandExo outputs in order to estimate the lightcurve precisions on each wavelength bin for both orders. We then use the relationship found in Figure \ref{fig:precision} in order to predict the 
%expected precision on the ``limb spectrum" at each wavelength bin{{. We follow a similar procedure to 
%find the covariance at each of those precisions between the limbs}}.

In order to showcase the importance of directly using transit lightcurves to constrain the limbs of 
exoplanets, and demonstrate that this indeed might be \textit{the most efficient avenue to constrain 
the transit spectrum of the limbs}, we use HAT-P-41b as a case-study{{, using both real 
\textit{HST} data and simulated \textit{JWST} observations}}. 

\begin{figure*}
\includegraphics[width=2.15\columnwidth]{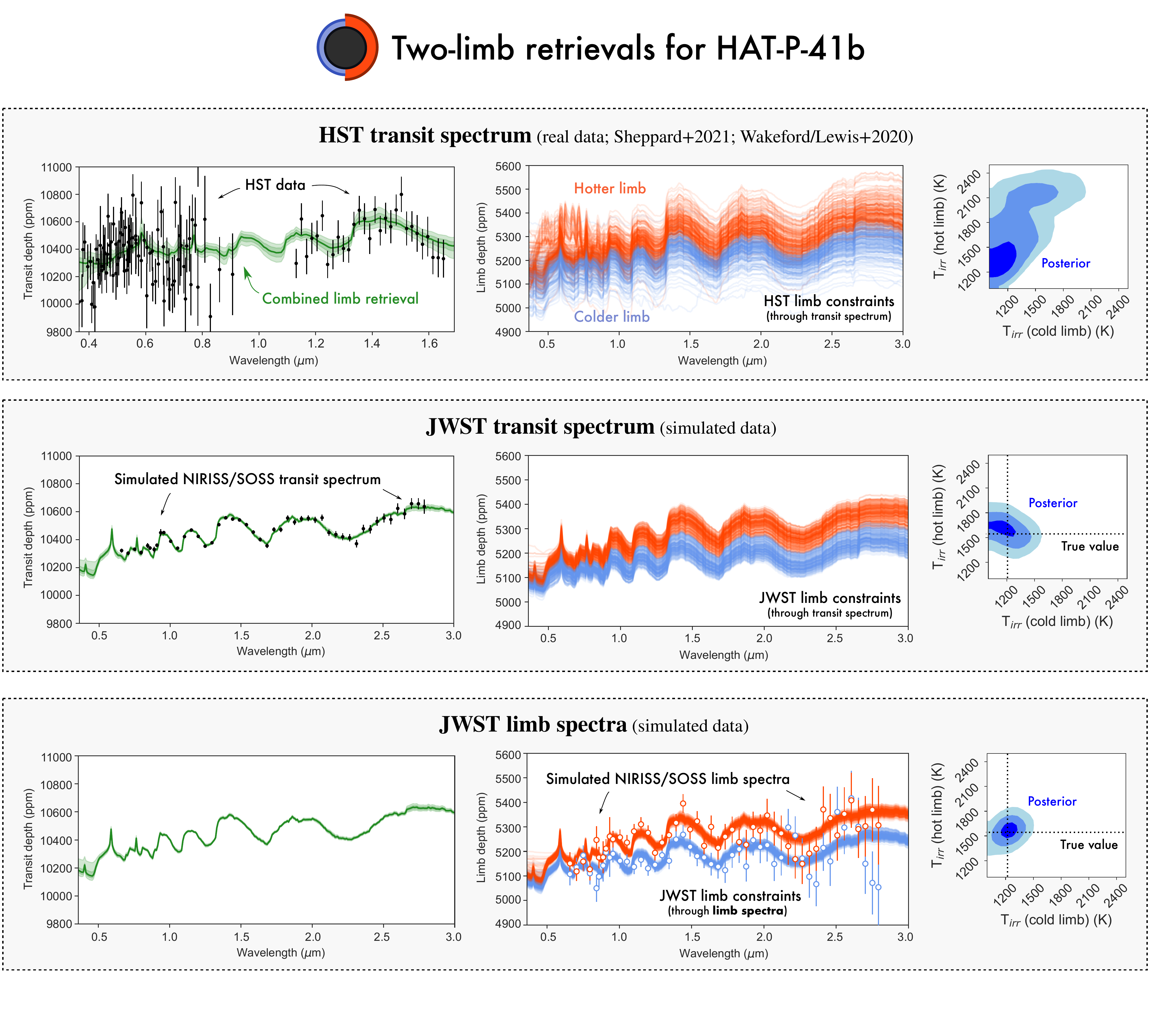}
    \caption{\textbf{Two-limb retrievals made on real \textit{HST} and simulated \textit{JWST} data.} (\textit{Top}) Observed transit spectrum by \textit{HST} STIS ($<1.0\ \mu m$) and \textit{HST} WFC3/IR ($>1.0\ \mu m$; black points with errorbars) of HAT-P-41b presented in \cite{w2020}, \cite{l2020} and \cite{s2020}, along with the 
    retrieved transit spectrum (left; green bands representing the 68\% and 95\% credibility intervals), 200 random 
    draws from the posterior retrieved limb models (middle) and posterior distribution of the temperatures of 
    those limbs (right). (\textit{Middle}) Same, but for two-limb retrievals made on a simulated \textit{JWST} 
    NIRISS/SOSS transit spectrum. (\textit{Bottom}) Same, but for a retrieval made on the \textit{limb spectra}, 
    extracted directly from simulated transit lightcurves. See text for details. \textbf{Main point:} Retrievals 
    made directly on limb spectra (bottom panel) constrain much better the limb models than retrievals made 
    on ``classic" transit spectra (top and middle panels).}
    \label{fig:limb-spectra}
\end{figure*}

\begin{figure*}
\includegraphics[width=2.15\columnwidth]{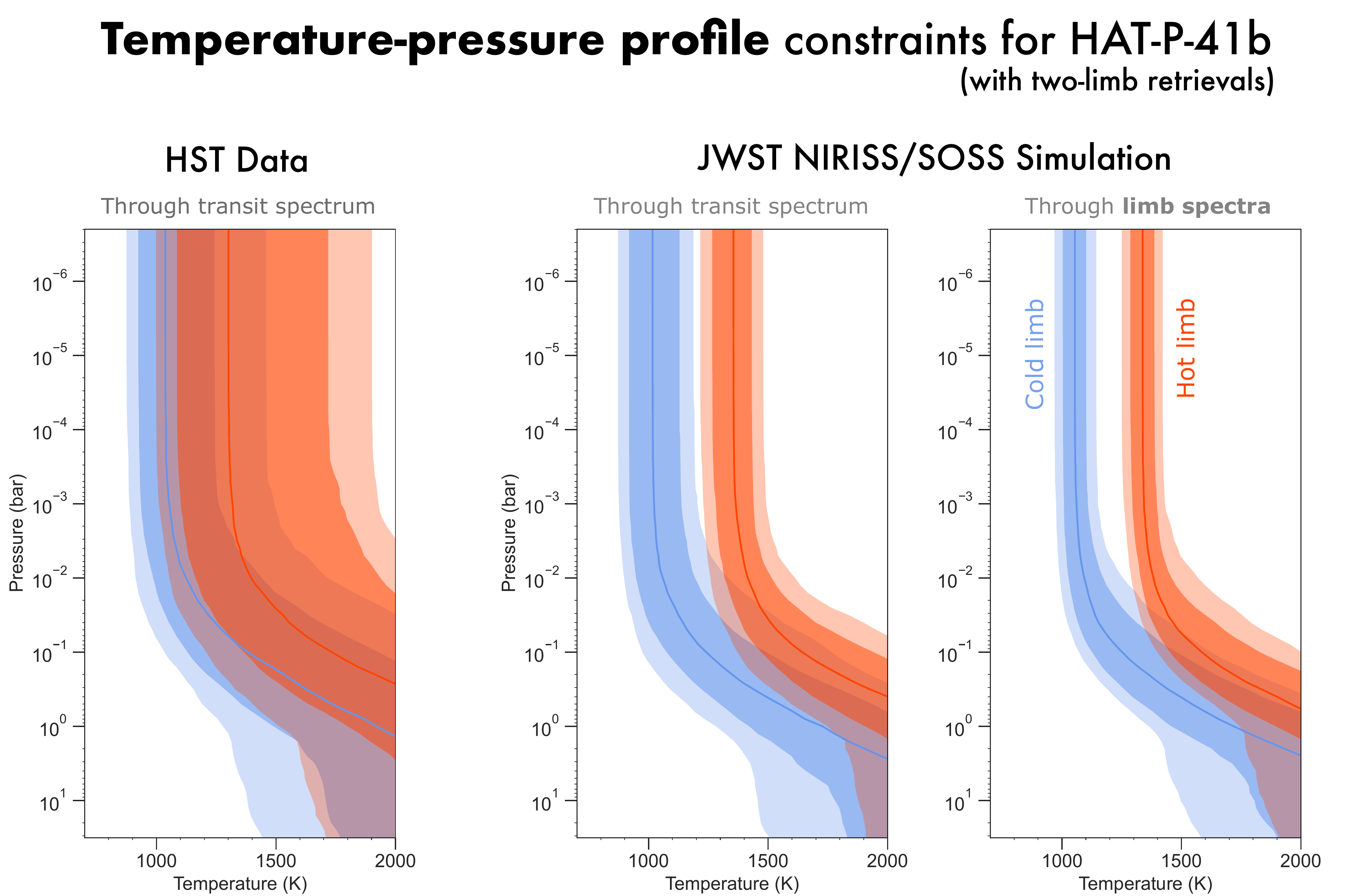}
    \caption{\textbf{Retrieved HAT-P-41b temperature-pressure profiles from two-limb retrievals}. (\textit{Left}) Profiles for the 
    cold (blue) and hot (red) limbs for the two-limb retrievals performed on \textit{HST}'s transit spectrum, show 
    in Figure 7 (top panel). (\textit{Middle}) Profiles for the same limbs and retrieval performed on the simulated 
    \textit{JWST}'s transit spectrum, shown in Figure 7 (middle panel). (\textit{Right}) Profiles for the same limbs as 
    for the middle panels, but performed on \textit{JWST}'s \textit{\textbf{limb}} spectra, shown in Figure 7 (lower panel). 
    \textbf{Main point:} limb retrievals extracted through limb spectra provide much more precise temperature-pressure profiles 
    for each limb at the pressures probed by transit spectroscopy ($\sim 1$ mbar).}
    \label{fig:pt-profiles}
\end{figure*}

{{We first analyze the {{\textit{HST}}} data presented in \cite{w2020}, \cite{l2020} 
and \cite{s2020}{{. In particular, we take the HST/WFC3 (1.1-1.6 $\mu$m), STIS G430L 
(0.3-0.5 $\mu$m) and G750L (0.5-0.9 $\mu$m) data from \cite{s2020}, and the WFC3 UVIS G280 
(0.2-0.8 $\mu$m) data from \cite{w2020}. In order to perform the atmospheric retrievals, we use 
the}} framework developed in \cite{macdonald:2020}, namely, one in which the 
\textit{transmission spectrum} is fit with a model using \textit{two} different 
atmospheric structures, each of which represents one of the limbs. To this end, we use and slightly modify the 
``chemically consistent" \texttt{CHIMERA}\footnote{\url{https://github.com/mrline/chimera}} atmospheric retrieval framework 
of \cite{chimera}, such that two transmission spectra are modeled at each iteration of the algorithm. 
Each of those has a different temperature and cloud structure, as well as different C/O ratios; the temperature 
at the top of the atmosphere is forced to be always colder for one limb than for the other in order to simulate 
energy redistribution processes that might be happening between the limbs. {{Chemical equilibrium is 
assumed for each limb}}. Once those 
are computed, each model transit depth is multiplied by $1/2$ in order to compute the spectra 
of each of the limbs, $\delta_i$. These are then added together to form the ``combined" 
transit depth that we compare against the \textit{HST} transit spectrum. We decide to leave the 
metallicity, the 10 bar radius, and the overall temperature-pressure profile shape (other than 
the temperature at the top at the atmosphere) as common parameters between the limbs, as we assume 
the \textit{HST} data would not be sensitive to those parameters. {{We also only use 
the data at wavelengths longer than 0.35 $\mu$m, as the publicly available opacities within \texttt{CHIMERA} only go down to this particular wavelength range. }}The retrieval is performed using 
nested sampling with the the \texttt{pymultinest} library \citep{PyMultiNest} which makes use of the MultiNest \citep{MultiNest} algorithm. The full set of priors used for 
our \texttt{CHIMERA} atmospheric retrievals in what follows are presented in Table 
\ref{tab:retpriors}. 

\begin{deluxetable}{lrl}[b!]
\tablecaption{Priors and parameters used for our Section \ref{sec:retrievals} \texttt{CHIMERA} 
atmospheric retrievals. These are composed by two limbs, one forced to be colder than the other. 
$U(a,b)$ below stands for a uniform distribution between $a$ and $b$. For detailed discussions on 
each parameter, see \cite{chimera} for the general framework and 
\cite{chimera2} for the implementation of the cloud model used in this work, which is that 
of \cite{AM}. \label{tab:retpriors}}
\tablecolumns{3}
\tablenum{2}
\tablewidth{1pt}
\tablehead{
\colhead{Parameter} &
\colhead{Prior} & 
\colhead{Comment}
}
\startdata
\multicolumn{3}{l}{\textit{Parameters individual to each limb}} \\
%\vspace{0.1cm}
\ \ $T_{\textnormal{irr}}$ (K) & $U(1000, 2500)$ & Stellar input at \\
 & & top of atmosphere.\\
\ \ $\log_{10} C/O$  & $U(-2, 0.3)$ & Log C/O ratio. \\
\ \ $\log_{10} K_{zz}$ (cm$^2$/s)  & $U(5, 11)$ & Log-eddy diffusion \\
& & coefficient.  \\
\ \ $ f_{\textnormal{sed}}$ & $U(0.5, 5)$ & Sedimentation efficiency. \\
\ \ $ \log_{10} P_{\textnormal{base}}$ (bar) & $U(-6, 1.5)$ & Cloud base pressure.  \\
\ \ $ \log_{10} f_{\textnormal{cond}}$ & $U(-15, -2)$ & Cloud condensate mixing\\
& & ratio at cloud base.  \\
%\multicolumn{3}{@{}l}{\makecell{Parameters common to both limbs}} \\
%\vspace{0.1cm}
\multicolumn{3}{l}{\textit{}} \\
\multicolumn{3}{l}{\textit{Parameters common to both limbs}} \\
\ \ [M/H] & $U(-2,3)$ & Atmospheric metallicity.\\
\ \ $f_{R}$ & $U(0.5,1.5)$ & Multiplicative factor to \\
& & 10 bar ``fiducial" radius.\\
\ \ $\log_{10} \kappa_{IR}$ (cm$^2$/g) & $U(-3,0)$ & T/P profile IR opacity.\\
\ \ $\log_{10} \gamma_{1}$ & $U(-3,0)$ & Log visual-to-IR opacity.\\
%r1\tablenotemark{a} & $U(0,1)$ & \rone \\
%r2\tablenotemark{a} & $U(0,1)$ & \rtwo \\
\enddata
%\tablenotetext{a}{These parameters correspond to the parametrization presented in \citet{kipping:2013} for sampling physically plausible combinations of the quadratic 
%limb-darkening coefficients. }
%\tablenotetext{b}{Time-averaged equilibrium temperature computed according to equation~16 of \citet{mendez:2017}}
\end{deluxetable}

The best-fit retrieval to the \textit{HST} data using this two-limb retrieval 
framework is presented on the top panel of Figure \ref{fig:limb-spectra}{{; the corresponding 
retrieved temperature-pressure profile is also presented in the left panel of Figure \ref{fig:pt-profiles}}}. 
Posterior credibility intervals 
for each parameter fit in the retrieval are presented in the second column of 
Table \ref{tab:retrieval-posteriors}. As can be seen in Figure \ref{fig:limb-spectra}, while 
the retrieved transit spectrum follows the data fairly well, the individual retrieved limb 
spectra shows a wide range of possible solutions, suggesting that they are fairly unconstrained 
by the data. Indeed, the overall posterior values for all parameters are largely 
indistinguishable between the limbs. For example, the temperatures at the top of the atmosphere are $1210^{+227}_{-132}$ K and $1522^{+519}_{-241}$ K for the ``cold" and ``hot" 
limbs, respectively, which are completely consistent with each other. The rest of the parameters, while 
fairly unconstrained by the data, all point towards a cloudy nature in HAT-P-41b's atmosphere; 
our solution suggest a relatively high-altitude cloud-deck solution, whose base is at about 10 
mbar --- a picture in good agreement with the {{forward modelling}} performed in 
\cite{l2020}{{. While the retrievals in that work preferred a near-UV opacity source 
shaping the spectrum instead of clouds, most of these retrievals were done fitting the 
atomic and molecular abundances directly --- without chemical constraints between them. 
The only chemical equilibrium retrieval performed in it needed a large temperature 
($>$ 2500 K) to be able to build up the necessary H$^-$ opacity to explain the data better. 
The chemical equilibrium retrieval performed here, while including H$^-$ opacity, is much 
more constrained in temperature as we didn't allow the possibility for the temperature at the 
limbs to exceed 2500 K. Nonetheless, while we see a possible second mode in the temperature 
posterior in Figure \ref{fig:limb-spectra} where the hot limb can reach temperatures of about 
2300 K, we see no buildup of posterior density above this range. While our retrieval could 
thus be interpreted as an alternative solution to the observed \textit{HST} data, we don't discuss 
this in length here as the objective of this experiment was to only obtain \textit{a} possible 
solution for a two-limb model given the \textit{HST} data, in order to showcase the importance of 
constraining limb spectra with upcoming high-precision facilities, as we will illustrate and 
quantify below}}.

The next experiment we ran consisted of taking the median of all the parameters found from our 
two-limb retrieval of the HAT-P-41b \textit{HST} data --- i.e., the medians presented in the second 
column of Table \ref{tab:retrieval-posteriors} --- and use the corresponding limb models implied by 
them to simulate a single \textit{JWST} NIRISS/SOSS transit observation using the \texttt{catwoman} transit 
model introduced in this work. As in the last sub-section, we used PandExo to estimate the per-integration 
lightcurve precision for this transit observation, with which we simulated 608, 
53-second integrations --- i.e., a 8.95-hour exposure, the recommended time-on-target 
for \textit{JWST} observations, obtained using equation \ref{tdwell}. To generate the transit models at each wavelength, 
we used the same transit parameters used by \cite{w2020} with the exception of the radius for each limb, 
for which we use the model described above. As for limb-darkening, we used the ExoCTK 
\citep{exoctk} limb-darkening tool\footnote{\url{https://exoctk.stsci.edu/limb_darkening}} to compute 
the limb-darkening coefficients of the quadratic law for orders 1 and 2 of NIRISS/SOSS. 

We tackled the analysis of these simulated lightcurves in two different ways, in 
order to showcase why performing inference on the limb spectra is the best way to 
optimally extract information about the limbs of exoplanets. The first consisted 
of extracting \textit{transit} spectra from them by simply fitting these 
lightcurves with \texttt{batman} lightcurve models, with the same procedure explained 
in Section \ref{sec:jwst-like}. Then, we performed two-limb retrievals on this 
transit spectrum using the same modified \texttt{CHIMERA} atmospheric retrievals 
described above and performed on the \textit{HST} data. The results of that 
experiment are presented in the middle panel of Figure \ref{fig:limb-spectra}{{, the 
retrieved pressure-temperature profiles are also shown in the middle column of Figure \ref{fig:pt-profiles}}}; 
posterior credibility intervals for each fitted parameter fit in the retrieval are presented in the third column of Table \ref{tab:retrieval-posteriors}. As can be seen, these two-limb retrievals constrain much better the different 
limbs than the \textit{HST} data, but still have a fairly large uncertainty, 
such that the retrieved limb spectra overlaps between the limbs at certain 
wavelengths. While the posteriors for most parameters are fairly wide (mainly due to 
the cloudy nature of this particular exoplanet), the temperature posterior 
distribution is much better constrained than the corresponding one for the \textit{HST} 
data: they have uncertainties that are about 1 to 2-fold smaller for the cold limb 
and about 2 to 5-fold smaller for the hot limb. 

Next, we performed the transit fitting using our \texttt{catwoman} model, again, using the 
same procedures as the ones outlined in Section \ref{sec:jwst-like}; this allowed us to 
extract \textit{limb spectra directly from the transit lightcurves}. In order to perform 
retrievals on these limb spectra, however, we had to slightly modify the log-likelihood of 
our retrievals because, as explained in the previous sub-section, for each wavelength bin the 
two limb depths are strongly correlated. We characterize the posterior distributions of the 
limb depths at each wavelength bin in our simulations as multivariate gaussians, {{which 
are described by}} covariance {{matrices}} $\Sigma_w$ and mean vectors 
$\vec{\mu}_w = [\delta_{1,w}, \delta_{2,w}]$ for each wavelength $w${{. The covariance matrices were estimated in the case of our fits by computing the 
sample covariance matrix of the posterior samples of the limb depths at each 
wavelength bin.}} With this, the log-likelihood we 
use in our retrievals is given by:
\begin{eqnarray*}
\ln \mathcal{L} = -\frac{1}{2} \sum_{w} 2\ln 2\pi + \ln |\Sigma_w| + \vec{r}_w^T \Sigma_{w}^{-1}\vec{r}_w
\end{eqnarray*}
with $\vec{r}_w = \vec{\mu}_w - \vec{m}_w$ being the vector of residuals, with $\vec{m}_w$ being the 
vector containing the model limb spectra at the given wavelength bin $w$. 

The results of performing the retrieval on the limb spectra directly using the 
framework described above are presented in the bottom panel of 
Figure \ref{fig:limb-spectra}{{, with the corresponding retrieved pressure-temperature profiles 
shown in the right column of Figure \ref{fig:pt-profiles}}}; posterior credibility intervals for each fitted parameter fit in the retrieval are presented in the fourth column of Table \ref{tab:retrieval-posteriors}. As can be observed, the constraints on the limb 
spectra are much better than the ones obtained from fitting the transit spectra. 
The corresponding limb spectral models are much better differentiated by our 
retrievals, allowing us to tightly constrain any features that might arise in them. 
The temperature posteriors for the limbs are also much better constrained; in 
particular, there is a 2 to 4-fold precision increase on the temperature of the 
cold limb, and 3 to 8-fold precision increase in the hot limb when compared 
against the \textit{HST} data. {{ In addition, the correlation between the retrieved 
limb temperatures is much smaller for this limb retrieval, as can be observed on the posterior 
distribution presented in Figure \ref{fig:limb-spectra}. This tighter constrain on the temperature can also 
be visually observed in Figure \ref{fig:pt-profiles}, where the much tighter constrain in the temperature-pressure 
profile of each limb is illustrated as well.}} These results showcase that extracting and analyzing 
limb spectra directly allows us to constrain much more tightly the properties of 
the limbs than methods that rely on extracting this information from ``classic" 
transit spectra {{{\footnote{The full set of posterior samples for all parameters along with the data 
used to perform the retrievals can be found in the following Github repository: \url{https://github.com/nespinoza/hat-p-41b-retrieval-posteriors}}.}}}
 }}
 
\begin{deluxetable*}{lrrr}[b!]
\tablecaption{Posterior credibility intervals (C.I.) of the two-limb retrieval made on real \textit{HST} data 
and simulated \textit{JWST/NIRISS} data of HAT-P-41b. C.I. below correspond to medians and 68\% credibility bands around it. For details on the definition for each parameter and priors, see Table \ref{tab:retpriors}. \label{tab:retrieval-posteriors}}
\tablecolumns{4}
\tablenum{3}
\tablewidth{1pt}
\tablehead{
\colhead{Parameter} &
\colhead{Posterior C.I.} & 
\colhead{Posterior C.I.} & 
\colhead{Posterior C.I.} \\
\colhead{} & 
\colhead{(transit spectrum, \textit{HST})} & 
\colhead{(transit spectrum, \textit{JWST}\tablenotemark{a})} & 
\colhead{(limb spectra, \textit{JWST}\tablenotemark{a})}
}
\startdata
\multicolumn{4}{l}{\textit{Parameters for the ``cold" limb}} \\
%\vspace{0.1cm}
\ \ $T_{\textnormal{irr}}$ (K) & $1210^{+227}_{-132}$ & $1197^{+130}_{-115}$ & $1237^{+59}_{-70}$ \\
\ \ $\log_{10} C/O$  & $-0.85^{+0.56}_{-0.65}$ & $-0.98^{+0.42}_{-0.57}$ & $-0.99^{+0.42}_{-0.56}$\\
\ \ $\log_{10} K_{zz}$ (cm$^2$/s)  & $7.9^{+1.7}_{-1.6}$ & $8.0^{+1.9}_{-1.9}$ & $7.7^{+1.9}_{-1.7}$\\
\ \ $ f_{\textnormal{sed}}$ & $3.4^{+1.6}_{-1.7}$ & $3.4^{+1.6}_{-1.7}$ & $3.3^{+1.8}_{-1.7}$\\
\ \ $ \log_{10} P_{\textnormal{base}}$ (bar) & $-2.1^{+1.9}_{-1.9}$ & $-1.9^{+2.1}_{-2.5}$ & $-1.8^{+2.1}_{-2.6}$  \\
\ \ $ \log_{10} f_{\textnormal{cond}}$ & $-7.0^{+3.1}_{-4.5}$ & $-9.3^{+3.6}_{-3.5}$ & $-9.2^{+3.9}_{-3.5}$\\
\multicolumn{4}{l}{\textit{}} \\
\multicolumn{4}{l}{\textit{Parameters for the ``hot" limb}} \\
%\vspace{0.1cm}
\ \ $T_{\textnormal{irr}}$ (K) & $1522^{+519}_{-241}$ & $1592^{+94}_{-108}$ & $1567^{+66}_{-76}$\\
\ \ $\log_{10} C/O$  & $-0.68^{+0.50}_{-0.78}$ & $-1.16^{+0.49}_{-0.52}$ & $-1.23^{+0.51}_{-0.48}$\\
\ \ $\log_{10} K_{zz}$ (cm$^2$/s)  & $8.0^{+1.7}_{-1.7}$ & $7.8^{+2.0}_{-1.8}$ & $7.8^{+2.1}_{-1.8}$\\
\ \ $ f_{\textnormal{sed}}$ & $3.1^{+1.6}_{-1.5}$ & $3.3^{+1.7}_{-1.6}$ & $3.3^{+1.7}_{-1.7}$\\
\ \ $ \log_{10} P_{\textnormal{base}}$ (bar) & $-2.5^{+2.0}_{-1.9}$ & $-1.9^{+2.1}_{-2.5}$ & $-1.9^{+2.1}_{-2.5}$ \\
\ \ $ \log_{10} f_{\textnormal{cond}}$ & $-8.0^{+3.8}_{-4.0}$ & $-9.1^{+3.7}_{-3.6}$ & $-9^{+3.8}_{-3.7}$\\
%\multicolumn{3}{@{}l}{\makecell{Parameters common to both limbs}} \\
%\vspace{0.1cm}
\multicolumn{4}{l}{\textit{}} \\
\multicolumn{4}{l}{\textit{Parameters common to both limbs}} \\
\ \ [M/H] & $2.06^{+0.24}_{-0.37}$ & $2.02^{+0.036}_{-0.041}$ & $2.022^{+0.037}_{-0.039}$\\
\ \ $f_{R}$ & $1.0276^{+0.0072}_{-0.0087}$ & $1.0273^{+0.0017}_{-0.0027}$ & $1.0276^{+0.0014}_{-0.0022}$\\
\ \ $\log_{10} \kappa_{IR}$ (cm$^2$/g) & $-2.23^{+0.51}_{-0.50}$ & $-1.92^{+0.75}_{-0.61}$ & $-2.32^{+0.57}_{-0.43}$\\
\ \ $\log_{10} \gamma_{1}$ & $-1.57^{+0.72}_{-0.81}$ & $-1.74^{+0.77}_{-0.75}$ & $-1.76^{+0.95}_{-0.80}$\\
%r1\tablenotemark{a} & $U(0,1)$ & \rone \\
%r2\tablenotemark{a} & $U(0,1)$ & \rtwo \\
\enddata
\tablenotetext{a}{The \textit{JWST} simulations had as underlying true values the medians from our \textit{HST} retrievals. }
%\tablenotetext{b}{Time-averaged equilibrium temperature computed according to equation~16 of \citet{mendez:2017}}
\end{deluxetable*}

{{Aside from impacting directly on inferences made with atmospheric retrievals, as has been shown above, 
we would like to highlight that extracting limb spectra directly from transit lightcurves 
has an additional, unique benefit: it opens up the discovery space for atmospheric features 
that might be individual to each of the limbs. This, in turn, has the benefit of 
allowing a direct comparison against Global Circulation Models (GCM). GCM 
modeling assumptions and implementation details have been actively driven 
by observations in the past few years. For instance, until recently clouds and 
hazes were typically added in a post-processing stage, not including feedback from 
these components in the modeling. Recent works such as those of \cite{rr:2019} 
and \cite{p:2020} have managed to implement clouds in their GCMs directly, 
showing in turn that this is critical to understand the overall cloud structure 
itself and heat redistribution in the planets under study, which has provided a 
much better match to observed phase-curve observations. We believe the study of limb spectra 
can open up a similar set of insights providing key advancements in the overall modeling of exoplanetary 
atmospheres.}} 

{{From an observational perspective, it is interesting to consider that }}observations aiming to extract these limb spectra can be much cheaper than, e.g., 
phase-curves. Whereas {{a single}} NIRISS/SOSS transit for HAT-P-41b, for instance, would amount to a total science time to 
a \textit{JWST} proposal of about {{8.95}} hours (using Equation \ref{tdwell}), a full phase-curve for this particular 
exoplanet requires over 65 hours. In this sense, extraction of limb spectra could serve as a good diagnostic as to 
what to expect in a phase-curve observation \textit{before} performing these expensive observations. Since for a fixed planetary and stellar radii the ingress/egress duration increases with the square-root of the semi-major axis (i.e., 
$\tau \propto \sqrt{a}$), this technique for detecting limb asymmetries might in turn be an excellent alternative avenue 
to studying morning and evenings of longer period planets where phase-curve signals are too small to be detectable in a 
reasonable amount of time.

\subsection{Timing variation biases due to asymmetric terminator depths}
\label{sec:timing}

In Section \ref{sec:detection} we discussed how, as predicted by \cite{vParis}, small changes in the time-of-transit center 
can give rise to equally good fits on symmetric models (such as the ones assumed by the \texttt{batman} package, for instance), 
even if the data is truly arising from an asymmetric transit model {{\citep[although this limitation can be bypassed 
if the aim is to retrieve limb spectra by simultaneously fitting transit lightcurves at different wavelengths; see, e.g.,][]{powell:2019}}}. It is important to note that this in turn can give 
rise to biases in transit times if a symmetric model is used when the data in fact comes from an asymmetric model such as 
the one modeled by the \texttt{catwoman} library. In our simulations, these can give rise to timing offsets of up to 
5 seconds, which is in turn within the timing precision that the \textit{TESS} mission is able to reach, and will be 
for sure within reach of the \textit{JWST} mission. Care must be taken, thus, when searching for small (second-level) 
timing offsets in these precise transit lightcurves in the search of, e.g., transit timing variations. 

\subsection{Limitations of this study}

It is important to note that throughout this work, we have assumed that the only alternative model to that of 
limb asymmetries is that of a symmetric limb in order to explain transit lightcurves asymmetries. However, there are other competing effects that might give rise to asymmetric lightcurves as 
well. For instance, known stellar effects such as gravity darkening \citep[see, e.g.,][and references therein]{ahlers} 
and yet-to-be uncovered effects/properties such as exoplanetary rings \citep[see, e.g.,][and references therein]{rein} can 
also give rise to asymmetric lightcurves. Performing a detailed model comparison study between these effects and the one 
studied here is out the scope of this work, but we warn researchers that proper care must be taken when aiming to claim 
the detection of asymmetric limbs in light of these possibilities. While effects like, e.g., gravity darkening are 
most likely known at good enough precisions in order to understand when a given transit lightcurve might be asymmetric 
due to this effect or at the very least to put limits on asymmetries generated by it, known unknowns such as 
exoplanetary rings might be more complicated to rule out. Perhaps the easiest way to constrain this would be through 
the wavelength-dependence of these asymmetries, which we hypothesise should be markedly different in the case 
of exoplanetary rings and those produced by opacity sources in an exoplanetary atmosphere. Still, it is 
important to be mindful of these alternative hypotheses when analyzing data on the search of these lightcurve asymmetries.

{{In addition to the above, the very validity of the framework developed here 
--- i.e., fitting transit lightcurves with a model of two ``stacked", rotated semi-circles --- remains 
to be put to the test with real data, and has plenty of room for improvement as we increase 
the precision of our measurements. For instance, our model assumes a sharp discontinuity at 
the poles, whereas some GCMs actually predict smooth transitions at them 
\citep[see, e.g., ][and references therein]{pluriel:2020}, even suggesting morning and evening 
terminators might be \textit{themselves} asymmetric. While a path in the 
right direction, our model is just an approximation to reality by construction, as those 
simplifications were the ones that allowed us to create a modeling scheme that is fast 
and efficient to apply to real transit lightcurve data. We expect that the detection of 
the signatures of mornings and evenings on actual, precise transit lightcurves 
from \textit{TESS} or \textit{JWST} could indeed motivate more flexible and accurate models 
for their shapes guided by GCM modeling (or the data themselves) that could expand on the 
simple modeling scheme discussed in this work. While these shapes might be complex enough that simple geometrical 
arguments like the ones made in this work would most likely not be easy to make, making 
the problems hard to parametrize, we are hopeful that good ideas might flourish in the 
exoplanet community to make this happen. There currently exists a continuum of lightcurve 
analysis methods ranging from few-number-of-parameter but constrained models like ours to 
very flexible but large-number-parameter models like the shadow imaging technique presented in 
\cite{sandford:2019}. We believe expanding our model in the direction discussed above lies 
in between those methodologies, and is thus bounded --- and therefore approachable. 
Developing this idea further, however, is outside the scope of this work.
}}

\section{Conclusions}
\label{sec:conclusions}

In this work, we have presented a detailed study on the observational prospects of directly detecting 
transit lightcurve asymmetries due to inhomogeneous exoplanetary limbs with current and 
near-future instrumentation. A semi-analytical framework was introduced in Section \ref{sec:thealgorithm} to fit transit lightcurves in 
order to extract the transit depths of the different limbs, a problem which is approximated as a pair of 
stacked semi-circles of different radii transiting a limb-darkened star following \cite{vParis}. Implemented in the 
\texttt{catwoman} python library \citep{jones}, this framework allows for the fast computation of these 
lightcurves, which are even able to model the rotation of the axis that joins the semi-circles, being thus 
able to characterize sky-projected planetary spin-orbit misalignmements in a complementary fashion to that 
allowed by the eclipse mapping technique for both lightcurves \citep{r7,w6} and radial-velocities \citep{n15}. 

A detailed feasibility study was presented in Section \ref{sec:detection} for detecting the effect with 
current existing facilities such as \textit{TESS} and near-future observatories such as \textit{JWST}. Even in a 
worst-case scenario of a planetary transit with a very small ingress/egress duration (which is the portion of the 
lightcurve that mainly constrains the limbs), the prospects for detecting the effect are very promising, even 
considering our ignorance on the angle that defines the sky-projected planetary spin-orbit misalignmement. If aiming at detecting the effect with only one transit, however, care must be taken as the time of transit center is highly 
degenerate with the limb asymmetries (i.e., a small shift in the time-of-transit can give rise to a similarly 
good fit to that of an asymmetric lightcurve due to inhomogeneous limbs). 

Finally, we showed in Section \ref{sec:discussion} how important the transit ingress/egress duration is for the 
detection of the effect. We used HAT-P-41b as a case-study to showcase the prospects for 
extracting the spectra of each of its limbs, and concluded that analyzing the lightcurves directly with the methods 
presented in this work might be one of the most efficient ways to obtain a global picture of each of the limbs, with \textit{JWST}-like precisions enabling the extraction of their spectra given the exquisite spectrophotometric precision 
the observatory will be able to achieve. 

We believe the promise of being able to characterize the limbs of exoplanets could play a pivotal role in our 
understanding of the 3-dimensional structure of exoplanets, and could provide observations that can inform current 
(e.g., GCM, transmission spectroscopy models) and future (e.g., phase-curves) models and observations aimed at the characterization of exoplanet atmospheres. The technique might, in turn, be a much less time-demanding technique to 
probe the 3D structure of longer period exoplanets, where phase-curves can be prohibitively expensive. Overall, we 
believe exploring the detectability of the effect in real transit lightcurves is critical to understand the limitations 
of the technique of transit spectroscopy when it comes to interpreting structural profiles such as temperature/pressue profiles 
and abundances in a 1-dimensional fashion \citep{macdonald:2020}. This, in turn, will be fundamental to make claims regarding 
formation mechanisms of these exoplanets based on the latter \citep{oberg,espinoza,mordasini}, and their overall 
dependance with planetary properties \citep[see, e.g.,][and references therein]{sing,wellbanks}.

\acknowledgments

{{We would like to thank the anonymous referee for their excellent feedback which 
significantly improved the presentation of our results.}}. We would also like to thank L. Carone and P. Molli\`ere for fruitful discussions on the theory of transmission 
spectroscopy and 3D modeling as applied to asymmetric limbs, as well as L. Kreidberg, T. Louden, D. Powell 
and N. Nikolov on fruitful discussions regarding constraining limb inhomogeneities directly from transit 
lightcurves. N.E. would like to thank the IAU and the Gruber foundation for the support to this research through 
the IAU-Gruber Fellowship, with which this work was started. We would also like to thank the MPIA Summer 
Program for the support to start this research. This research made use of the open source Python package exoctk, the Exoplanet Characterization Toolkit \citep{exoctk}.

%%%%%%%%%%%%%%%%%%%%%%%%%%%%%%%%%%%%%%%%%%%%%%%%%%
%%%%%%%%%%%%%%%%%%%%%%%%%%%%%%%%%%%%%%%%%%%%%%%%%%
%%%%%%%%%%%%%%%%%%%%%%%%%%%%%%%%%%%%%%%%%%%%%%%%%%
%%%%%%%%%%%%%%%%%%%%%%%%%%%%%%%%%%%%%%%%%%%%%%%%%%
\appendix

%%%%%%%%%%%%%%%%%%%%%%%%%%%%%%%%%%%%%%%%%%%%%%%%%%
%%%%%%%%%%%%%%%%%%%%%%%%%%%%%%%%%%%%%%%%%%%%%%%%%%
%%%%%%%%%%%%%%%%%%%%%%%%%%%%%%%%%%%%%%%%%%%%%%%%%%
%%%%%%%%%%%%%%%%%%%%%%%%%%%%%%%%%%%%%%%%%%%%%%%%%%
\section{Deriving $\Delta A$}
\label{sec:deltaA}
In order to derive the decrement of flux due to the transit of a pair of stacked semi-circles given by 
equation \ref{eq:theproblem}, we must find $\Delta A$, 
the inter-sectional area between the stacked 
semi-circles and the iso-intensity band depicted in 
Figure \ref{fig:geometry}. As already noted by 
\cite{batman} in the case of a circle, this area is 
simply the inter-sectional area of the stacked semi-circles with the circle of radius $x_i$, $A(x_i,R_{p,1},R_{p,2},d,\varphi)$, minus the same inter-sectional area but with the circle of radius $x_{i-1}$, $A(x_{i-1},R_{p,1},R_{p,2},d,\varphi)$, i.e.,
\begin{eqnarray*}
\Delta A = A(x_i,R_{p,1},R_{p,2},d,\varphi) - A(x_{i-1},R_{p,1},R_{p,2},d,\varphi).
\end{eqnarray*}
This implies that to find $\Delta A$ one has to 
first find a general form for the inter-sectional 
area between the stacked semi-circles and a circle. These stacked semi-circles, in turn, are 
composed of two semi-circles, and thus the problem reduces to calculating the area of the intersection between a circle and two (rotated) 
\textit{semi-circles} with a common center: one of radius $R_{p,1}$ and 
another of radius $R_{p,2}$, but rotated by 180 degrees. Given a general formula 
for such intersection, $A_S(R,R_S,d,\theta)$, 
where $R$ is the radius of the circle, 
$R_S$ the radius of the semi-circle, $d$ the distance 
between the center of the circle and the semi-circle 
and $\theta$ the rotation angle of the semi-circle 
with respect to $d$, then
\begin{eqnarray*}
A(x,R_{p,1},R_{p,2},d,\varphi) = A_S(x,R_{p,1},d,\varphi) + A_S(x,R_{p,2},d,\varphi + \pi).
\end{eqnarray*}
If we find a general form for $A_S(\cdot)$, then we solve the problem. We tackle this problem in the next 
sub-section.

\subsection{Intersection area between a circle and a semi-circle}
\label{sec:semicircle}
Although the case of calculating the intersection area between two circles can be obtained via elemental trigonometry, the problem of calculating the intersection area between a circle and a (rotated) \textit{semi}-circle is not, in general, as straightforward. 

\begin{figure*}
\includegraphics[width=0.5\columnwidth]{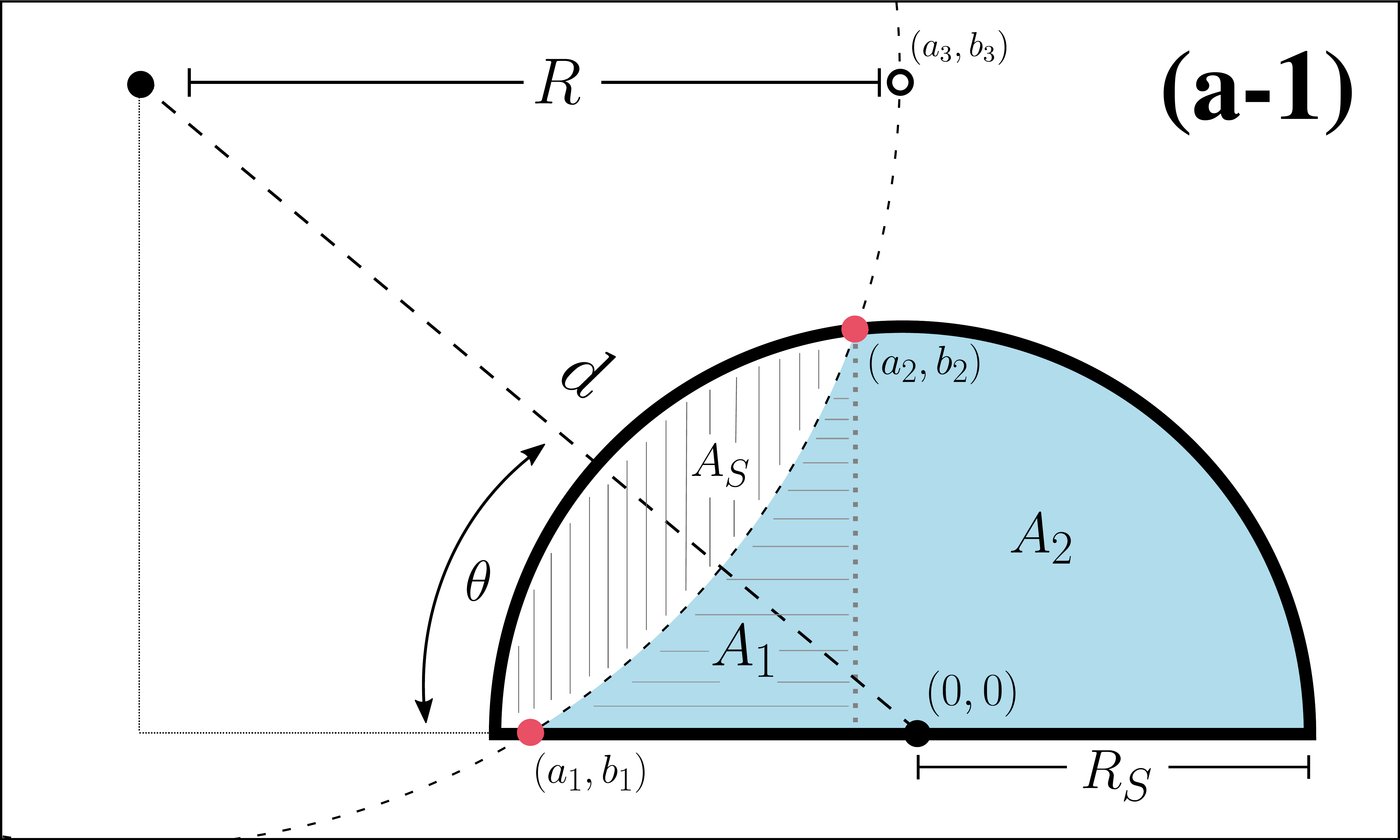}
\includegraphics[width=0.5\columnwidth]{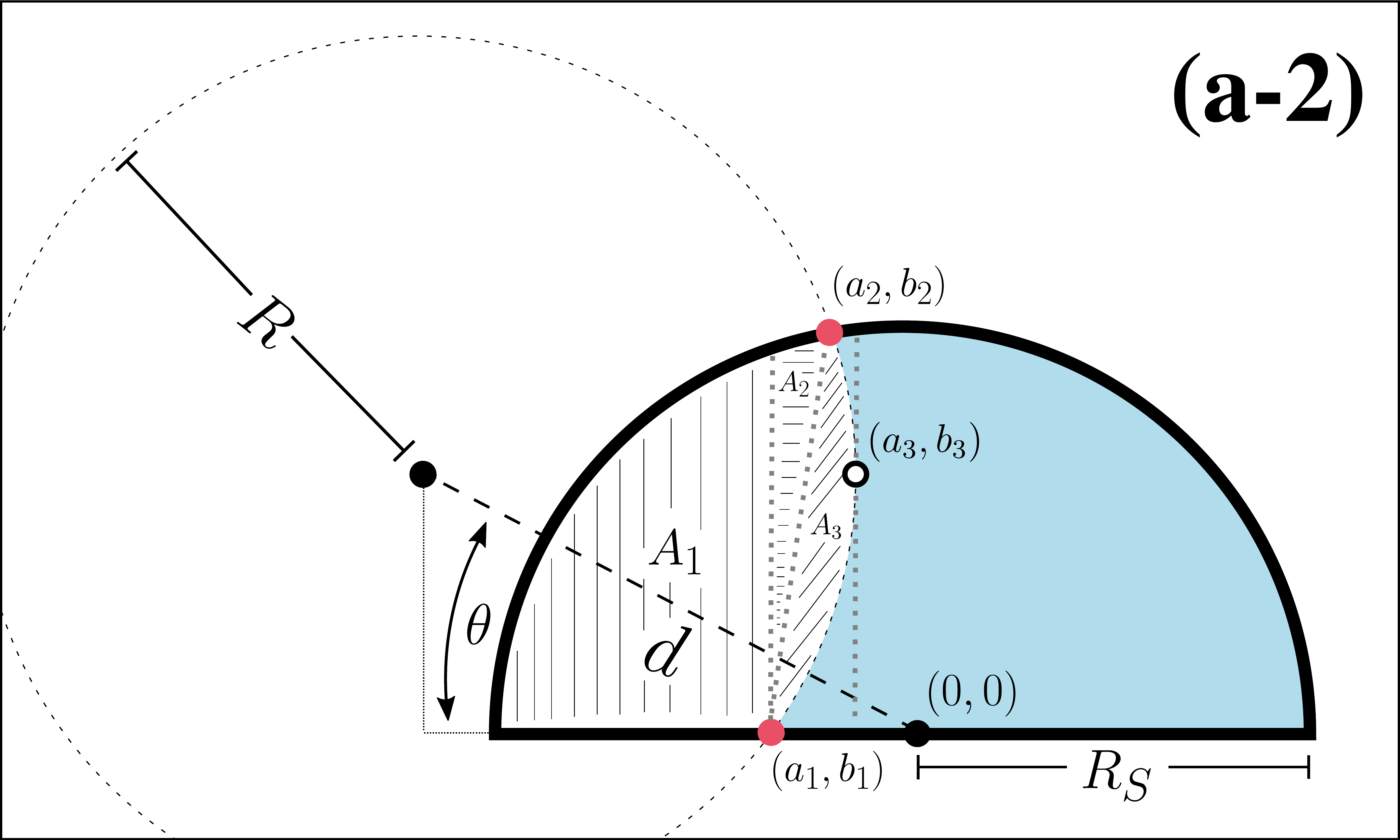}
\includegraphics[width=0.5\columnwidth]{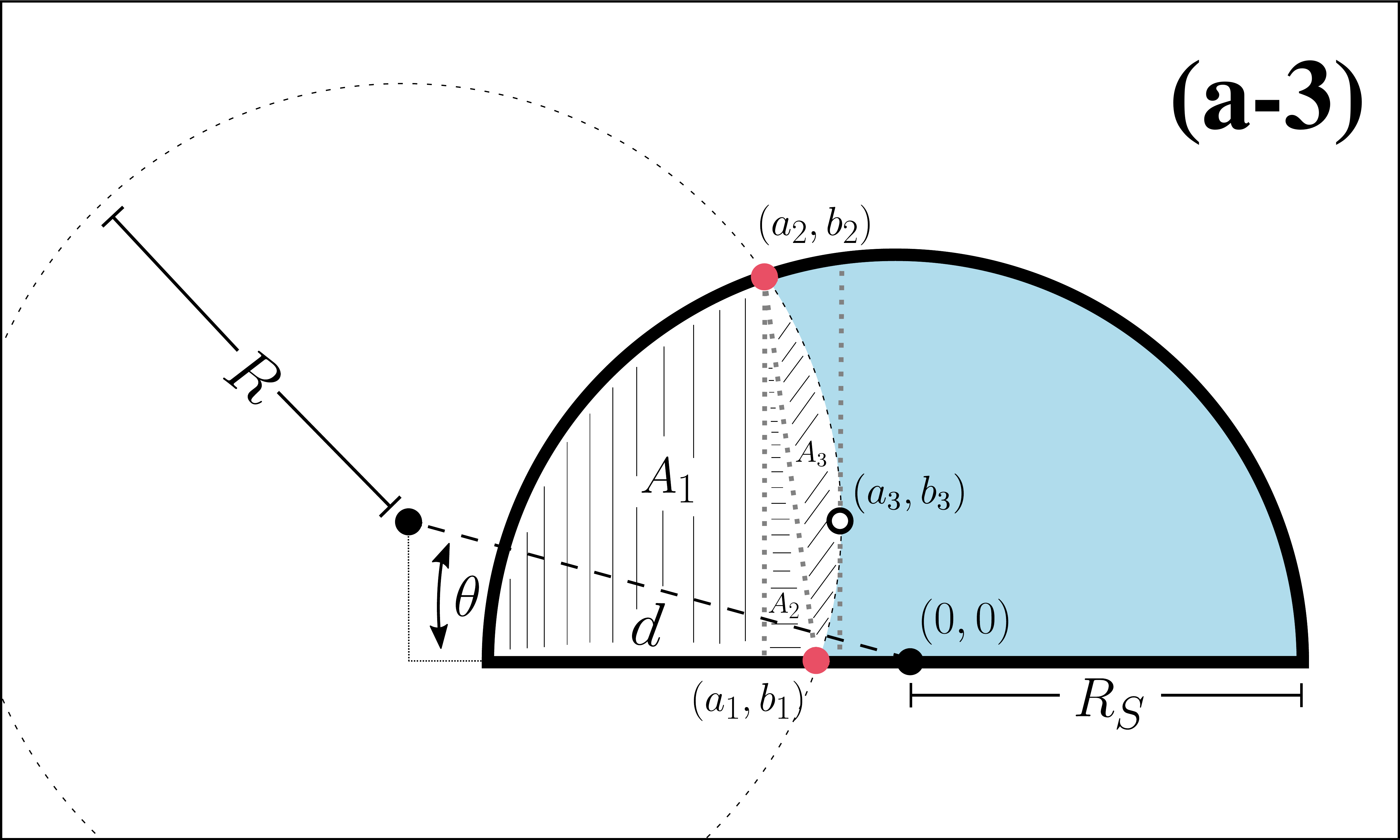}
\includegraphics[width=0.5\columnwidth]{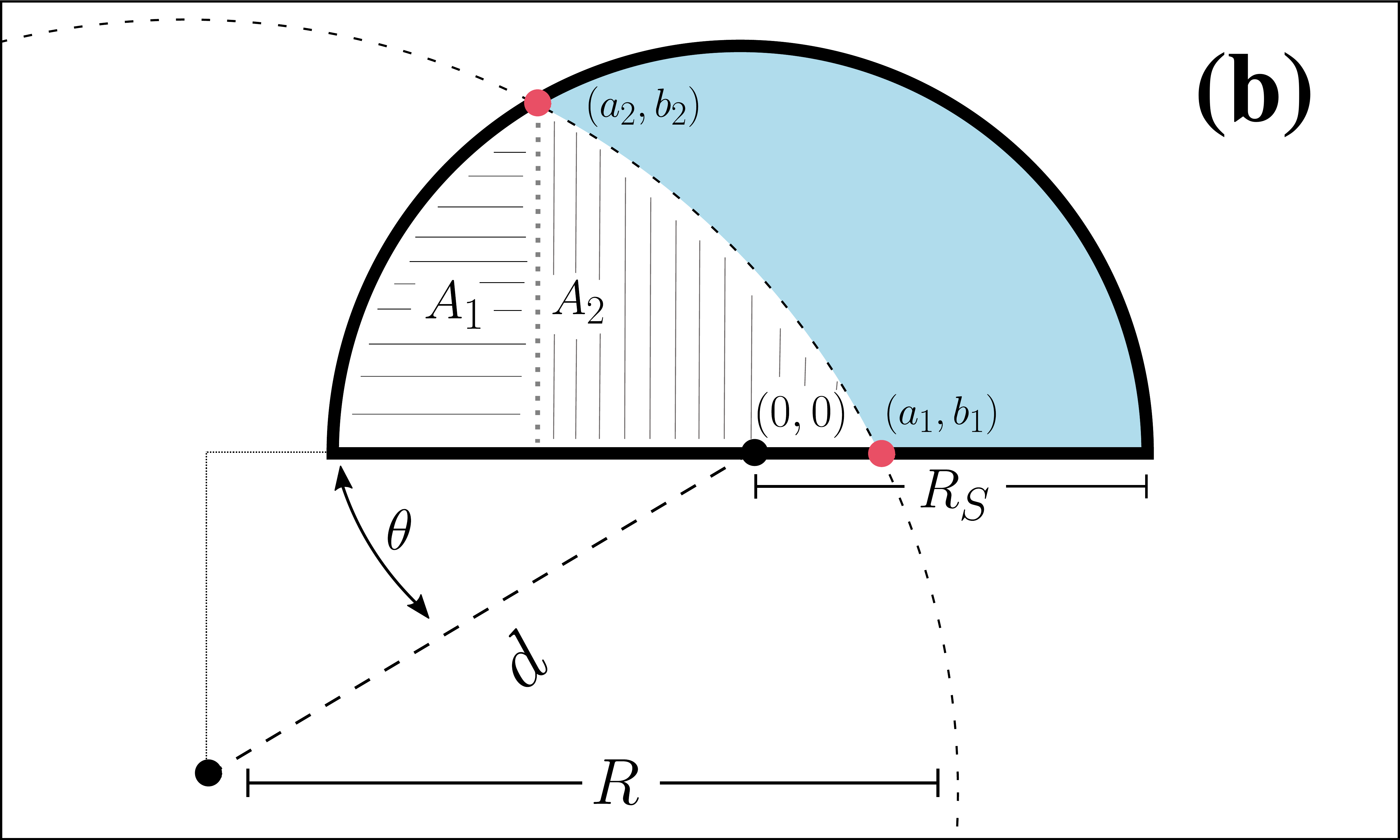}
\includegraphics[width=0.5\columnwidth]{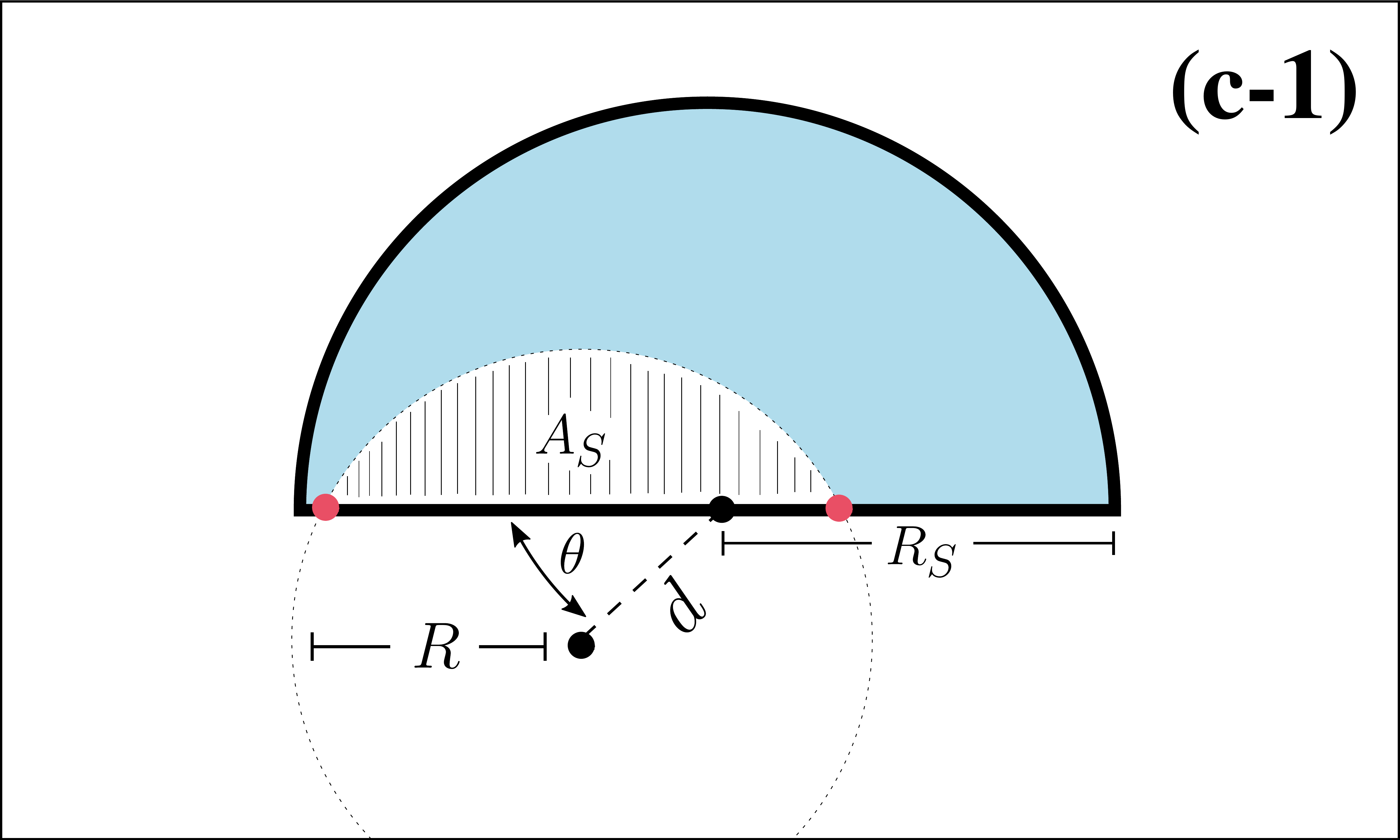}
\includegraphics[width=0.5\columnwidth]{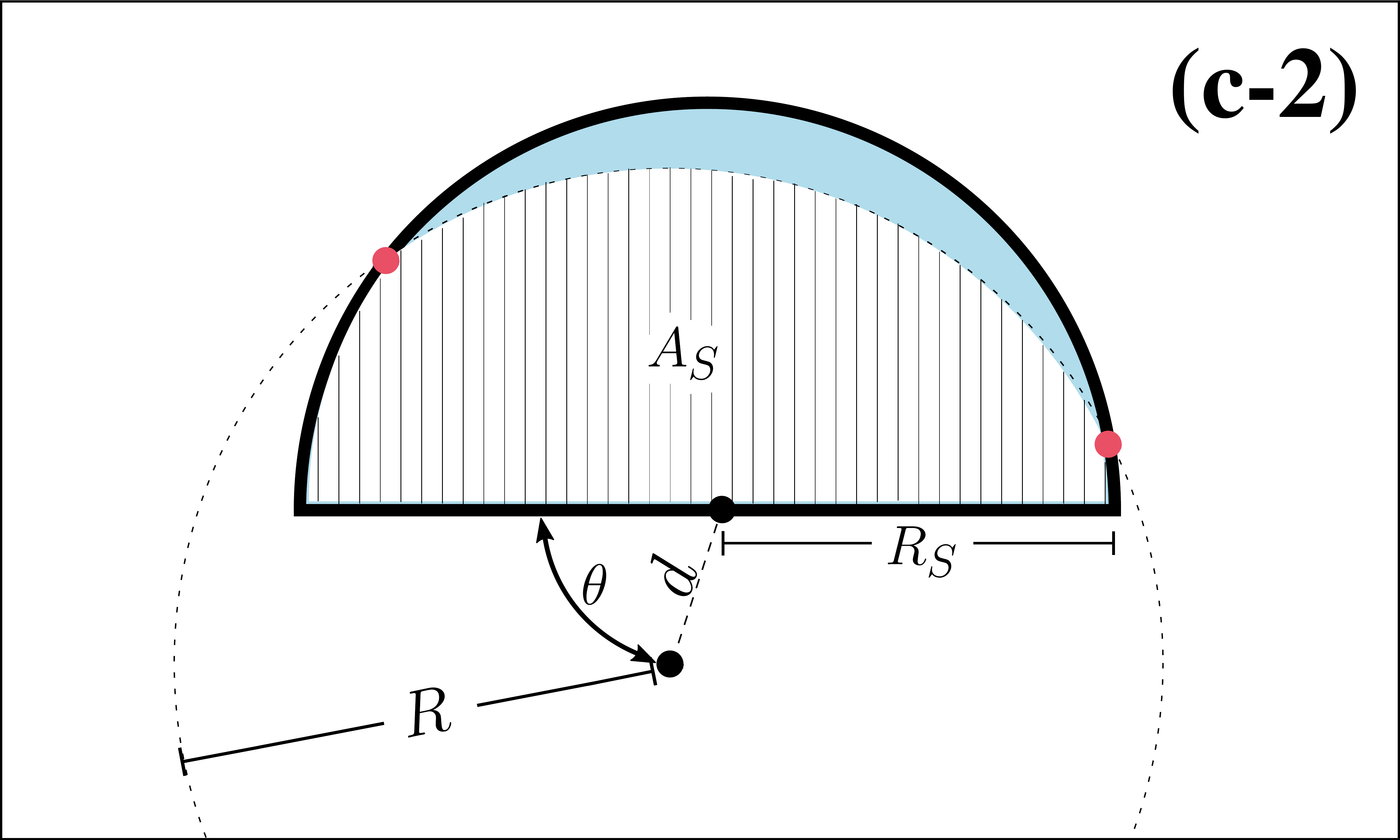}
\includegraphics[width=0.5\columnwidth]{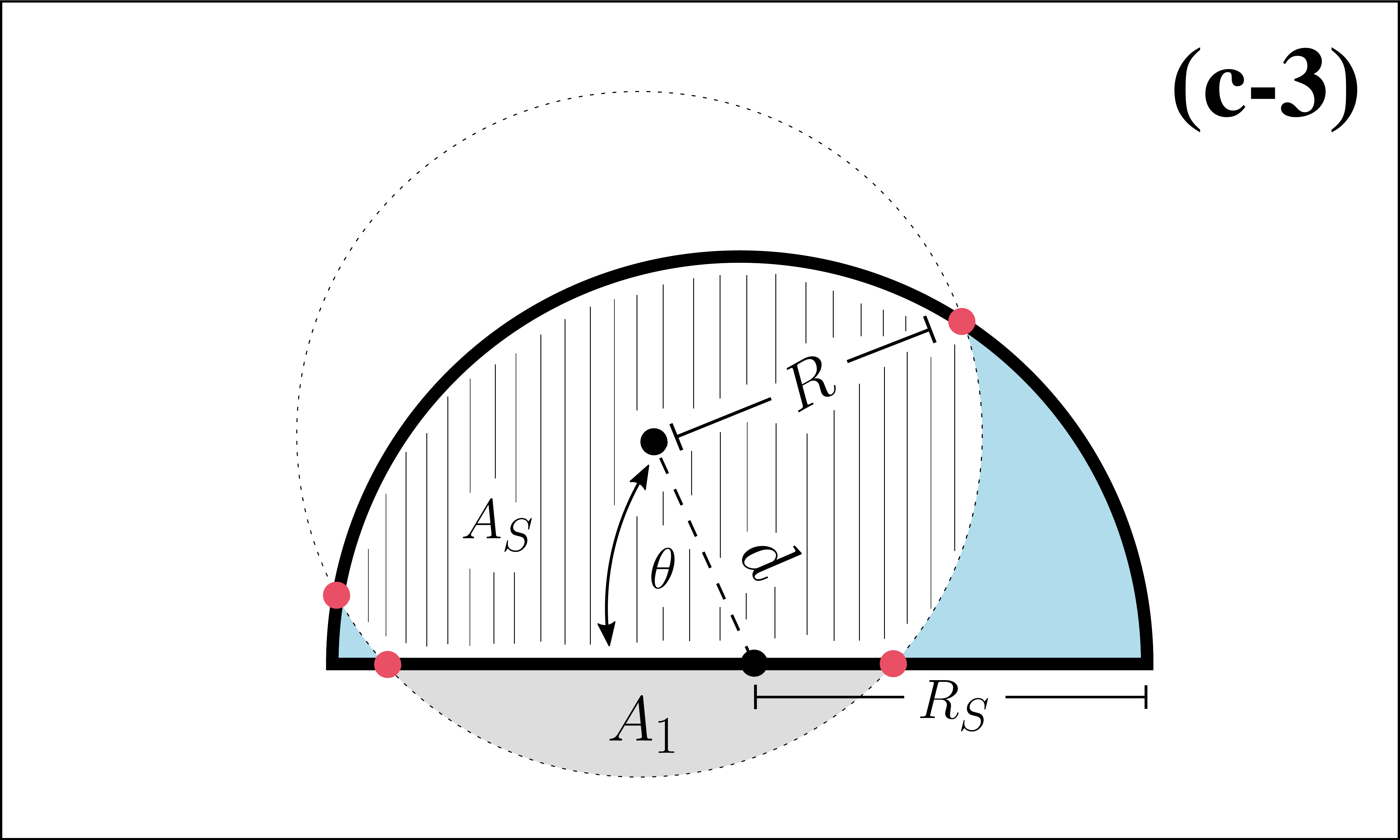}
    \caption{Transformed geometry of the problem --- we chose to rotate the circle of radius $R$ around the semi-circle of radius $R_S$ by an angle $\theta$. This problem is the same as having the semi-circle rotated with respect to the line that joins the centers (of length $d$) by an angle $\theta$. The general problem can, in turn, be divided in three cases: (a) when the center of the circle is above the base of the semi-circle (divided, in turn 
    in sub-cases a-1, a-2 and a-3), (b) when the center of 
    the circle is below the base of the semi-circle and the intersection points are not 
    both touching the base or the edge of the semi-circle and (c) when there are two (c-1) or more (c-2) simultaneous intersection points with either at the edge or at the base of the 
    semi-circle; this latter case can be solved using basic trigonometry. In all cases, 
    the intersection points between the semi-circle and the circle are indicated by red dots. 
    In all cases, the area of interest is the white dashed one inside the semi-circles.}
    \label{fig:geometry2}
\end{figure*}

We first note that the problem of finding the intersection area of a circle of radius $R$ 
and a semi-circle of radius $R_S$ rotated by an angle $\theta$ with respect to the 
line that joins the centers of length $d$ is the same problem as the intersection 
area of a semi-circle and a \textit{circle rotated by an angle $\theta$ with respect 
to the line that joins the centers}. This symmetry argument allows us to put the horizontal 
axis of this problem in the base of the semi-circle, simplifying the notation of the 
problem. Without loss of generality, we put the origin in the center of the base of the 
semi-circle. This transformed geometry of the problem is shown in Figure \ref{fig:geometry2}; 
here the white dashed area inside the semi-circle is the area of interest (i.e., the one that 
leads to $A_S(R,R_S,d,\theta)$).

As is evident in Figure \ref{fig:geometry2}, there are three different cases (a, b and c) 
we have to take care of in order to find a general formula for $A_S(R,R_S,d,\theta)$:
\begin{itemize}
    \item Case (a), divided into sub-cases (a-1), (a-2) and (a-3), deals with 
    the problem in which the circle is rotated such that it lies 
    above the semi-circle. If we identify the coordinates of the center of the circle 
    as $(x_0,y_0)=(-d\cos \theta, d\sin \theta)$, case a) deals with the problem in 
    which $\theta > 0$ and, thus, $d\sin \theta > 0$. Here, the intersection points between 
    the circle and the semi-circle have coordinates $(a_1,b_1)$ and $(a_2,b_2)$. The 
    geometry depicted in Figure \ref{fig:geometry2} for this case implies that $b_1=0$. 
    This is because for $b_1>0$, the problem is the same as 
    the intersection of two circles (one of radius $R$ and another of radius $R_S$), which 
    has a known analytical solution \citep[see, e.g., ][]{batman}. Here the area of interest,
    $A_S$, is given by the area of the semi-circle ($\pi R_S^2/2$) minus $A_1+A_2$ for case 
    (a-1) and by $A_1+A_2+A_3$ for cases (a-2) and (a-3). The different sub-cases depend, in 
    turn, on the location of the intersection points and the position of $(a_3,b_3)$, the 
    position of the maximum extension of the circle in the x-axis. 

    \item Case (b) deals with the problem in which the circle is rotated such that it lies 
    \textit{below} the semi-circle, i.e., where $\theta < 0$ and, thus, $d\sin \theta < 0$. 
    In addition, this case handles only problems in which one intersection of the circle with 
    the semi-circle is in its base and the other is with the upper part of the semi-circle. Once 
    again, the intersection points between the circle and the semi-circle have 
    coordinates $(a_1,b_1)$ and $(a_2,b_2)$. In this case, however, $b_2 = 0$; the cases in 
    which $b_2\neq 0$ (i.e., when the right-most intersection is on the upper part of the 
    semi-circle) and in which $b_2 = 0$ \textit{and} $b_1 = 0$ (i.e., in which the 
    left-most intersection is also on the base of the semi-circle) is taken care of by case c). 
    The area of interest for case (b) is $A_S = A_1 + A_2$. 
    
    \item Finally, case (c), divided into sub-cases (c-1), (c-2) and (c-3), deals with the problem in which the circle is either rotated above or below the semi-circle, but where there are two 
    intersections with both either in the base (c-1) or in the upper part of the semi-circle (c-2) or 
    four intersection points (c-3) between the circle and the semi-circle. In this case, the 
    area of interest, $A_S$ can be calculated directly via basic trigonometry and thus we 
    don't identify here the intersection points by coordinates but by the red points in order 
    to guide the reader.
\end{itemize}

Cases (a), (b) and (c) defined above will all be calculated assuming that the center of the 
circle is to the left of the semi-circle. The reason for doing this is that the problem 
has reflective symmetry with respect to the line that goes through the center of the 
semi-circle and that is perpendicular to its base. As such, for $0 < \theta < \pi/2$ we 
have that $A_S(R,R_S,d,\theta) = A_S(R,R_S,d,\pi - \theta)$, whereas for 
$-\pi/2 < \theta < 0$, $A_S(R,R_S,d,\theta) = A_S(R,R_S,d,\theta - \pi)$. In what follows, we solve each of the cases separately.

\subsubsection{Case (a)}
Before looking at the integrals that lead to the intersection area between the circle and 
the semi-circle in this case, let us obtain the expressions for the intersection points 
$(a_1,b_1)$ and $(a_2,b_2)$. To do this, we first note that the point where the circle 
intersects with the line $y = 0$ (which is the line that passes through the base of the 
semi-circle) is given by 
\begin{eqnarray}
x_\textnormal{int} = -d\cos \theta \pm \sqrt{R^2 - d^2 \sin^2\theta}. 
\label{eq:a-1A0}
\end{eqnarray}
If the discriminant of this expression is negative (or zero), which in this case implies 
that $|d\sin \theta| \geq R$, then there is no (or one) intersection of the circle with 
the x-axis. In this case, the intersectional area can either be zero if $d \geq R+R_S$ or 
equal to solving the problem of the intersection between two circles of radius $R$ and 
$R_S$ in the case in which $d < R+R_S$. If the latter case is true, in turn, the intersectional 
area reduces to $\pi R^2$ when $d+R < R_S$ (i.e., when the circle is inside the semi-circle).

If $|d\sin \theta| < R$, then we have two solutions for the intersection points of the 
circle and the x-axis, $x_\textnormal{int}$. The cases in which both solutions are inside 
the semi-circle, i.e., when $|x_\textnormal{int}|\leq R_S$, will be handled in case (c), 
as depicted in Figure \ref{fig:geometry2}. If both solutions are \textit{outside} the semi-circle, 
i.e., when $|x_\textnormal{int}|> R_S$, there are two possibilities. If $d \geq R+R_S$, 
the intersection area is zero --- in this case, the circle is always to the left of the 
semi-circle. If, on the other hand, $d < R+R_S$, then because the intersections with 
the x-axis happen outside the semi-circle, there are two options: either the circle intersects twice with 
the upper part of the semi-circle (a case that will be solved in case (c-2)) or the circle covers all of the 
semi-circle, in which case the intersectional area is $\pi R_S^2/2$. If $\theta > 0$, then this leaves us, finally, with 
the problem we will solve for case (a) (is this is not true then case (b) will apply), in which $x_\textnormal{int}$ has both one solution outside 
and another one inside the semi-circle. Because here we are dealing with the cases in which the 
center of the circle is always to the left of the center of the semi-circle, this implies that 
the intersection point inside the semi-circle will always be the right-most intersection point, i.e., 
the solution of $x_\textnormal{int}$ with the positive sign. In the notation of Figure 
\ref{fig:geometry2}, this gives the intersection points
\begin{eqnarray*}
a_1 &=& -d\cos \theta + \sqrt{R^2 - d^2 \sin^2\theta},\\
b_1 &=& 0.
\end{eqnarray*}
The intersection point between the upper part of the semi-circle and 
the circle, $(a_2,b_2)$, is obtained by simply equating the equation of the circle ($(x+d\cos{\theta})^2 +(y-d\sin{\theta})^2 = R^2$) with the equation of the semi-circle, taking it as a full circle ($x^2 + y^2 = R_S^2$) to begin with. This yields
\begin{eqnarray}
\label{eq:a-1A0b}
b_2 &=& -A\sin \theta + \cos \theta \sqrt{R_S^2-A^2},\\
\label{eq:a-1A0c}
a_2 &=& b_2 \tan \theta + \frac{A}{\cos \theta},
\end{eqnarray}
where
\begin{eqnarray*}
A &=& \frac{R^2-R_S^2-d^2}{2d}.
\end{eqnarray*}
As we are not dealing with two intersecting circles, but an intersecting circle and semi-circle, we have chosen the largest $b_2$ which will give rise to the $a_2$, and therefore the point of intersection, that is on the semi-circle. %check this makes sense
%Here two solutions arise (one with a positive sign in equation (\ref{eq:a-1A0b}) and 
%negative sign in $B$ and one with negative sign in equation (\ref{eq:a-1A0b}) and 
%positive sign in $B$) as these intersection points consider the possibility that 
%the semi-circle might have two intersections with the circle of radius $R$ (as, e.g., 
%in the case where the center of this circle is directly above the center of 
%the semi-circle). Because of this, to find the coordinates of the intersection 
%point when there is only one intersection, one has to see which of the two solutions 
%gives rise to a real solution for $b_2$; thus, it suffices to check which of the 
%two solutions satisfies $R_S^2\cos^2\theta \geq AB$. 

Finally, an important set of coordinates to define are the ones for $(a_3,b_3)$. As 
illustrated in Figure \ref{fig:geometry2}, these are the coordinates of the maximum 
value attained in the x-axis by the circle. The coordinates for this point are, 
evidently,
\begin{eqnarray*}
a_3 &=& -d\cos \theta +R, \\
b_3 &=& d\sin \theta.
\end{eqnarray*}

First, we take on \textbf{case (a-1)}. This 
case occurs when the conditions for case (a) 
are met and when the point $(a_3,b_3)$ is 
\textit{outside} the semi-circle, which 
in turn implies in this case that $b_3\geq b_2$. To solve it, the strategy to obtain $A_S$ is to compute analytic 
solutions to the areas $A_1$ and $A_2$ depicted in Figure \ref{fig:geometry2}, 
and then subtract these to $\pi R_S^2/2$. First, $A_1$ is simply the area under the curve of 
the circle from $x=a_1$ to $x=a_2$. Because in this case $a_1 < a_2 < a_3$, 
we are going to integrate the lower part of the circle; this implies 
the equation of the (in this case semi) circle is simply 
\begin{eqnarray*}
y = -\sqrt{R^2 - (x + d\cos \theta)^2} + d\sin |\theta|.
\end{eqnarray*}
Integrating this from $x=a_1$ to $x=a_2$ yields
\begin{eqnarray}
\label{eq:a-1A1}
A_1 = \frac{R^2}{2}\Delta f + d\sin |\theta| (a_2 - a_1),
\end{eqnarray}
where $\Delta f = f(a_1)-f(a_2)$, with
\begin{eqnarray*}
f(x) = \left(\frac{x + d\cos \theta}{R^2}\right)\sqrt{R^2 - (x + d\cos \theta)^2} + \arcsin \left(\frac{x+d\cos \theta}{R}\right).
\end{eqnarray*}
% Note that the second integral is really |cos(t)|cos(t), but it goes to cos^2(t) because
% the integral limits are arcsines, and thus bounded by -1 to 1, range within which 
% cos (t) is always positive (i.e., -pi/2 < -1 < t < 1 < pi/2).
%where to solve the intergral we made the change of variables 
%$x = R\sin(t) - d\cos(\theta)$, and used the trigonometric identity $\cos^2(t) = (1+\cos %2t)/2$. Here:
%\begin{eqnarray*}
%a_1' &=& \arcsin\left(\frac{a_1 + d\cos \theta}{R}\right),\\
%a_2' &=& \arcsin\left(\frac{a_2 + d\cos \theta}{R}\right).
%\end{eqnarray*}
%(2) $\pi R_S^2/2$ in the case in which $d < R+R_S$ and the circle covers the semi-circle 
%completely or (3) . For this latter case to happen, the circle of 
%radius $R$ has to be equal or larger than the distance from the center of the circle to 
%the right-most corner of the semi-circle (which is the largest possible distance from the 
%center of the circle to the semi-circle), which by the law of cosines gives the condition 
%$R \geq \sqrt{d^2 +R_S^2 -2dR_S\sin \theta}$.
Next, we work on obtaining area $A_2$. This is simply the area under the curve of the 
semi-circle, whose equation is $y = \sqrt{R_S^2 - x^2}$. Integrating this from $x=a_2$ to 
$x = R_S$ yields
\begin{eqnarray}
\label{eq:a-1A2}
A_2 = \frac{R_S^2}{2}\left[\frac{\pi}{2} - h(a_2)\right],
\end{eqnarray}
where
\begin{eqnarray*}
h(x) = \arcsin{\left(\frac{x}{R_S}\right)} + \frac{x}{R_S}\sqrt{1 - \frac{x^2}{R_S^2}}.
\end{eqnarray*}
%\begin{eqnarray*}
%g(x) = \arccos{\left(\frac{x}{R_S}\right)}-\frac{x}{R_S}\sqrt{1 - \frac{x^2}{R_S^2}},
%\end{eqnarray*}
Thus using the definitions for $A_1$ given in equation (\ref{eq:a-1A1}) and for $A_2$ given in 
equation (\ref{eq:a-1A2}), area $A_S$ is given in this case by
\begin{eqnarray*}
\boxed{A_S = \frac{\pi R_S^2}{2}-\frac{R^2}{2}\Delta f - d\sin |\theta| (a_2 - a_1) - \frac{R_S^2}{2}\left[\frac{\pi}{2} - h(a_2)\right]}.
\end{eqnarray*}
Now, we take on \textbf{case (a-2)}. In this case $b_3<b_2$, however, $a_2>a_1$. In this case, the 
area of interest is $A_S = A_1 + A_2 + A_3$, as 
depicted in Figure \ref{fig:geometry2}. First, 
area $A_1$ in this case is the area of the 
semi-circle from $x= -R_S$ to $x=a_1$. Integrating once again 
the equation of the semi-circle 
($y = \sqrt{R_S^2 - x^2}$) in this range 
one obtains
\begin{eqnarray}
\label{eq:a-2A1}
A_1 = \frac{R_S^2}{2}\left[h(a_1) + \frac{\pi}{2}\right].%\frac{R_S^2}{2}h(a_1).
\end{eqnarray}
Area $A_2$ in this case can be calculated as the area under the 
same semi-circle between $x=a_1$ and $x=a_2$ 
\textit{minus} $b_2(a_2-a_1)/2$, which is the 
area of the triangle formed between the 
points $(a_1,b_1)$, $(a_2,b_2)$ and $(a_2,0)$. 
Integrating the semi-circle between 
$x=a_1$ and $x=a_2$ and subtracting $b_2(a_2-a_1)/2$, we obtain
\begin{eqnarray}
\label{eq:a-2A2}
A_2 = \frac{R_S^2}{2}\left[h(a_2) - h(a_1)\right] - \frac{b_2(a_2-a_1)}{2}.
\end{eqnarray}
Finally, area $A_3$ reduces to obtaining the segment of a circle generated by drawing a chord 
between points $(a_1,b_1)$ and $(a_2,b_2)$. To this end, we ought to know the angle $\alpha$ (in radians) these points make with respect to 
the center of the circle. This can be easily 
obtained by the Law of Cosines to give
\begin{eqnarray*}
\alpha = \arccos\left(1 - \frac{(a_2-a_1)^2 + b_2^2}{2R^2}\right).
\end{eqnarray*}
With this, the area of the segment $A_3$ 
is thus, simply
\begin{eqnarray}
\label{eq:a-2A3}
A_3 = \frac{R^2}{2}\left(\alpha - \sin \alpha\right).
\end{eqnarray}
Finally, then, using the definition for 
$A_1$ in equation (\ref{eq:a-2A1}), 
for $A_2$ in equation (\ref{eq:a-2A2}) and 
for $A_3$ in equation (\ref{eq:a-2A3}), 
we get in this case 
\begin{eqnarray*}
\boxed{A_S = \frac{R_S^2}{2}\left[h(a_2) + \frac{\pi}{2}\right]- \frac{b_2(a_2-a_1)}{2} + \frac{R^2}{2}\left(\alpha - \sin \alpha\right)}.
\end{eqnarray*}
Finally, we solve \textbf{case (a-3)}. 
In this case again $b_3<b_2$, however, $a_1>a_2$. Here, equation 
(\ref{eq:a-2A1}) also applies for 
$A_1$, but the upper limit of the integral is in this case $a_2$ 
instead of $a_1$. This implies that in this case 
\begin{eqnarray}
\label{eq:a-3A1}
A_1 = \frac{R_S^2}{2}\left[h(a_2) + \frac{\pi}{2}\right].
\end{eqnarray}
To obtain area $A_2$ in this case, we note that here this 
is simply the area of the triangle formed by the points with 
coordinates $(a_1,b_1)$, $(a_2,b_2)$ and $(a_2,0)$. In this 
case, thus, 
\begin{eqnarray}
\label{eq:a-3A2}
A_2 = \frac{b_2 (a_1 - a_2)}{2}.
\end{eqnarray}
Finally, to obtain $A_3$ we use equation (\ref{eq:a-2A3}) which 
also applies for this case. Using then the definition for 
$A_1$ in equation (\ref{eq:a-3A1}), for $A_2$ in equation 
(\ref{eq:a-3A2}) and for $A_3$ in equation (\ref{eq:a-2A3}), 
we get in this case 
\begin{eqnarray*}
\boxed{A_S = \frac{R_S^2}{2}\left[h(a_2) + \frac{\pi}{2}\right] + \frac{b_2 (a_1 - a_2)}{2} + \frac{R^2}{2}\left(\alpha - \sin \alpha\right)}.
\end{eqnarray*}

\subsubsection{Case (b)}
\textbf{Case (b)} is similar in many ways to case (a), with the only difference that now the coordinates of the center of the circle change to $(-d\cos \theta, -d\sin |\theta|)$, and thus some functions and integration ranges change signs. In this case, the intersection points of the circle with the x-axis are the same as the ones given in equation (\ref{eq:a-1A0}), and thus all of the discussion given at the beginning of the past sub-section also applies for case (b). In particular, the intersection points $(a_1,b_1)$ 
and $(a_2,b_2)$ derived for case (a) 
are the same for this case.  

In this case, the area of interest is the sum of area $A_1$ and $A_2$. The former is the integral of the 
semi-circle circle from $x=-R_S$ to $x=a_2$, which 
is an integral which was already found in equation (\ref{eq:a-3A1}). As for area $A_2$, this is the integral of the upper part of the circle of 
radius $R$, i.e., of the function
\begin{eqnarray*}
y = \sqrt{R^2 - (x + d\cos \theta)^2} - d\sin |\theta|.
\end{eqnarray*}
However, the integral of this from $x=a_2$ to $x=a_1$ is exactly the same integral calculated in case (a-1), 
whose result is on equation (\ref{eq:a-1A1}), because 
the integrand there was the same integrand that we have here 
but multiplied by -1, and the limits of integration there were reversed with respect to the 
ones we have here (i.e., they went from $a_1$ to $a_2$) --- because inverting the limits of 
integration is the same as calculating the integral multiplied 
by -1, both effects cancel out. Thus, area $A_2$ in our case 
is area $A_1$ in case (a-1). Thus, for case (b), we have that 
$A_S = A_1 + A_2$, i.e.,
\begin{eqnarray*}
\boxed{A_S = \frac{R_S^2}{2}\left[h(a_2)+\frac{\pi}{2}\right]+\frac{R^2}{2}\Delta f + d\sin |\theta| (a_2 - a_1)}.
\end{eqnarray*}

\subsubsection{Case (c)}
Case (c) focuses on when $|d\sin \theta|<R$, i.e. there are two solutions for the intersection points of the circle with the line $y=0$.

More specifically, \textbf{case (c-1)} occurs also when $|x_\textnormal{int}|\leq R_{S}$, i.e. when the two solutions for the intersection points are inside the semi-circle and when the part of the circle above the intersection points is completely enclosed within the semi-circle (see Figure \ref{fig:geometry2}). This can be quantitatively described by theoretically 'extending' the semi-circle into a full circle of radius $R_S$. The coordinates of intersection of these two circles (setting the center of the circle of radius $R_S$ at the origin) can be found by substituting $x^2 +y^2 =R_S^2$ into $(x+d\cos{\theta})^2+(y+d\sin{\theta})^2=R^2$ to give the y coordinates:
\begin{eqnarray*}
y=-A\sin{\theta}\pm \cos{\theta}\sqrt{R_S^2-A^2},
\end{eqnarray*}
where
\begin{eqnarray*}
A=\frac{R^2-R^2_S-d^2}{2d}.
\end{eqnarray*}
This is similar to equations \ref{eq:a-1A0b} and \ref{eq:a-1A0c} except the position of the circles relative to the origin have been changed slightly. 

Therefore case (c-1) applies when $|x_\textnormal{int}|\leq R_{S}$ \textit{and either}:
\begin{itemize}
    \item There is no solution for the intersection of the two circles. This will occur when $A^2>R_S^2$. 
    \item Both of the y-coordinate solutions are real and negative, i.e. when $-A\sin\theta \pm \cos{\theta}\sqrt{R_S^2-A^2}<0$.
\end{itemize}

\textbf{Case (c-2)} occurs when $|x_\textnormal{int}|> R_S$ and when there are two intersection points on the curved edge of the semi-circle. For this case, the y-coordinates of intersection between the circle $R$ and the full circle of radius $R_S$ must be positive. Using the same equations as above, this is when $-A\sin\theta \pm \cos{\theta}\sqrt{R_S^2-A^2}\geq0$, where $A$ is defined as in case (c-1).

\textbf{Case (c-3)} occurs when $|x_\textnormal{int}|\leq R_S$ and when there are two further intersection points on the curved edge of the semi-circle, making a total of four intersections points. 
%Due to the subtle change in the definition of $\theta$ (see Figure \ref{fig:geometry2}), the equation for the y-coordinates of intersection are now 
%\begin{eqnarray*}
%y=-A\sin{\theta}\pm \cos{\theta}\sqrt{R_S^2-A^2}. 
%\end{eqnarray*}
Therefore case (c-3) is when $|x_\textnormal{int}|\leq R_S$ and $-A\sin\theta \pm \cos{\theta}\sqrt{R_S^2-A^2}\geq0$. 
\\ \\ \\
\textbf{To solve case (c-1)}, the points of intersection in the base of the semi-circle can be obtained via equation \ref{eq:a-1A0}:
\begin{eqnarray*}
x^\pm_\textnormal{int} = -d\cos \theta \pm \sqrt{R^2 - d^2 \sin^2\theta}. 
\end{eqnarray*}
The problem is then just calculating the area of the segment $A_S$ which is a well-known geometric problem with the solution 
\begin{eqnarray*}
\boxed{A_S=R^2\arccos{\left(\frac{y}{R}\right)}-xy,}
\end{eqnarray*}
where 
\begin{eqnarray*}
x&=&(1/2)(x^+_{\textnormal{int}}+|x^-_{\textnormal{int}}|), \\
&=&\sqrt{R^2-d^2\sin^2\theta}, \\
y&=&-d\sin{\theta}.
\end{eqnarray*}

\textbf{To solve case (c-2)}, the problem can be set up by first theoretically 'extending' the semi-circle into a full circle of radius $R_S$ and calculating the area of intersection of the two circles using the equations described in \cite{batman}:
\begin{eqnarray}
\label{eq:c-2A1}
A_{\textnormal{int}}=R^2\arccos{u}+R_S^2\arccos{v}-(1/2)\sqrt{w},
\end{eqnarray}
where
\begin{eqnarray*}
u &=& (d^2 + R^2-R_S^2)/(2dR), \\
v &=& (d^2 + R_S^2 - R^2)/(2dR_S), \\
w &=& (-d+R+R_S)(d+R-R_S)(d-R+R_S)(d+R+R_S).
\end{eqnarray*}

Using $A_\textnormal{int}$, one can find $A_S$ by subtracting half the area of the 'extended' circle of radius $R_S$ to yield
\begin{eqnarray*}
\boxed{A_S=A_\textnormal{int}-\pi R_S^2/2.}
\end{eqnarray*}

%$A_{\textnormal{int}}$ can be used to calculate the area marked $B$ in Figure \ref{fig:geometry2} by subtracting $A_{\textnormal{int}}$ from the area of a circle of radius $R_S$, $\pi R_S^2$, to give
%\begin{eqnarray*}
%B=\pi R_S^2 - (R^2\arccos{u}+R_S^2\arccos{v}-(1/2)\sqrt{w}).
%\end{eqnarray*}

%Finally, the intersection area between the circle and the semi-circle ($A_S$) can be obtained by subtracting area $B$ from the total area of the semi-circle, $\pi R_S^2/2$ to yield
%\begin{eqnarray*}
%\boxed{A_S=-\pi R_S^2/2 + %R^2\arccos{u}+R_S^2\arccos{v}-(1/2)\sqrt{w}.}
%\end{eqnarray*}

\textbf{To solve case (c-3)}, the problem needs to be split up into two parts. The first part involves finding the area of intersection, $A_\textnormal{int}$, of the circle and the semi-circle 'extended' into a full circle of radius R using the same method from part (c-2) and the second involves finding area $A_1$. As can be seen from the Figure \ref{fig:geometry2}, once these two areas are found, it is simply a matter of subtracting $A_1$ from $A_\textnormal{int}$ to find $A_S$.

%Firstly, to find $A_1$, a similar approach to case (c-2) is taken, where the semi-circle is 'extended' into a full circle, and the intersectional area between the two circles is calculated using equation \ref{eq:c-2A1}. $A_1$ can then be found by subtracting $A_{\textnormal{int}}$ from the area of the circle of radius $R$, $\pi R^2$:
%\begin{eqnarray*}
%A_1=\pi R^2 - (R^2\arccos{u}+R_S^2\arccos{v}-(1/2)\sqrt{w})
%\end{eqnarray*}
%where $u$, $v$ and $w$ are defined the same as for case (c-2).

To find $A_1$ is a very similar problem to case (c-1), the points of intersection along the base of the semi-circle can be found using equation \ref{eq:a-1A0} and then the problem reduces to that of finding the area of a segment, which is $A_1$ in this case. Following a similar method for case (c-1),
\begin{eqnarray*}
A_1=R^2\arccos{\left(\frac{-y}{R}\right)}+xy,
\end{eqnarray*}
where $x$ and $y$ are defined the same as for case (c-1), however note change in sign of $y$, due to the change of orientation of the shapes.
As mentioned, $A_\textnormal{int}$ is the same as in case (c-2) and is described in equation \ref{eq:c-2A1}.

Therefore the total intersection area $A_S$, is given by
\begin{eqnarray*}
\boxed{A_S=A_\textnormal{int}-A_1}
\end{eqnarray*}

\subsection{Deriving $\theta$ }
\label{sec:theta}
In this paper we have defined $\theta$ as the angle between the base of the semi-circle and the line that extends from the base of the semi-circle to the center of the circle (of length $d$). Importantly, $\theta$ is defined as \textit{positive} when extending clockwise, assuming the center of the circle is to the left of the semi-circle which, as explained in section \ref{sec:semicircle}, due to the symmetry of the problem, can always be achieved by flipping the frame of reference. 

$\theta$ is calculated from parameters that correspond to how the system physically appears in the sky. {To explain, we place the center of the star at the origin of a 2-dimensional Cartesian coordinate system. Due to the symmetry of the star, the planet is assumed to move from left to right (horizontally) across it so that the dY/dX gradient of the direction of motion of the planet at $t_0$ (inferior conjuction) is 0 (planet moves along the X-direction only). This coordinate system is \textit{different} from the frame of reference used in section \ref{sec:semicircle} when the frame is rotated so that the center of the base of the semi-circle is at the origin and the base lies along the y=0 axis.

$\theta$ therefore depends on the angle the semi-circle is rotated through with respect to the X-direction in the 2D system defined above, $\phi$, the impact parameter, $b$ and whether the semi-circle is originally to the left or the right of the center of the circle. The latter is characterised by the time, $t$ and the time of inferior conjunction, $t_0$ and by assuming the planet moves from left to right. It is important to note that $\phi$ will be static if the planet moves in an almost straight line across the star (along the X-direction only). However due to orbital mechanics, the orbital path may appear curved and as such the planet will rotate slightly as it crosses the star. This rotation will perturb $\phi$ in such a way as discussed in \ref{sec:changeinphi}.}

$\phi$ is defined in this paper between the range $-\pi/2$ to $\pi/2$, from the X-axis to the base of the \textit{top} semi-circle, where a positive angle is defined going in the anticlockwise direction. Once this is used to find the $\theta$ for $R_{p,1}$, to obtain the $\theta$ for $R_{p,2}$, one simply has to multiply $\theta$ by -1, due to the change in direction.

Table \ref{table:theta} shows how $\theta$ is calculated for the different values of $\phi$, $b$ and $t$ for $R_{p,1}$.
\begin{center}
 \begin{tabular}{| c c c| c |} 
 \hline
 $t$ & $b$ & $\phi$ & $\theta$ \\ [0.5ex] 
 \hline\hline
 $\leq t_0$ & positive & $ < \arccos{b/d}$ & $-\phi - \arcsin{b/d}$ \\ 
  &  & $ \geq \arccos{b/d}$ & $- \pi +\phi + \arcsin{b/d}$ \\
 \hline
 $\leq t_0$ & negative & $\leq -\arccos{|b|/d}$ & $ \pi +\phi - \arcsin{|b|/d} $ \\ 
  &  & $ > -\arccos{|b|/d} $ & $-\phi + \arcsin{|b|/d}$ \\
  \hline
 $> t_0$ & positive & $ < -\arccos{b/d}$ & $-\pi-\phi + \arcsin{b/d}$ \\ 
  &  & $\geq -\arccos{b/d} $ & $\phi - \arcsin{b/d}$ \\
 \hline
 $> t_0$ & negative & $ \leq \arccos{|b|/d}$ & $\phi + \arcsin{|b|/d}$ \\ 
  &  & $ > \arccos{|b|/d} $ & $\pi-\phi - \arcsin{|b|/d} $ \\
 \hline
\end{tabular}
\label{table:theta}
\end{center}

\subsection{Change in $\phi$ and the impact parameter of the semi-circle as a function of phase due to orbital mechanics}
\label{sec:changeinphi}
{Using the same coordinate system as described in \ref{sec:theta}, if the planet were to move in a straight line across the star, then the planet's impact parameter, $b$, would stay constant at the value of the impact parameter defined at $t_{0}$ (the time of inferior conjuction).} From \cite{seager} this is defined as 
\begin{eqnarray}
b = \frac{a\cos{i}}{R_{star}}\bigg(\frac{1-e^2}{1-e\sin{\varpi}}\bigg),
\label{eq:b}
\end{eqnarray}
where $a$ is the semi-major axis, $i$ is the inclination of the orbit, $e$ is the eccentricity, $\varpi$ is the longitude of periastron and $R_{\textnormal{star}}$ is the radius of the star. 
{However due to the orbital motion of the planet around the star, from the perspective of our X-Y system, the orbital path of the planet is curved across the sky, and so the impact parameter becomes a function of time.} For the most accurate model, $b$ used in the equations in \ref{table:theta} should be adjusted to 
\begin{eqnarray}
b = -r\sin{(\omega+f)}\cos{i},
\label{eq:b}
\end{eqnarray}
where 
\begin{eqnarray}
r &=& \frac{a(1-e^2)}{1+e\cos{f}},\\
\cos{f} &=& \frac{\cos{E} - e}{1- e\cos{E}},\\
E -e\sin{E}&=& \frac{2\pi}{T}(t - t_0).
\label{eq:b}
\end{eqnarray}
$T$ is the orbital period of the planet, $t$ is the current time of interest and $t_0$ is the time of periastron passage. The last equation can be solved numerically using a Newton-Raphson method.

{Furthermore, due to the change in gradient of the planet's orbit with respect to this coordinate system, the angle between the base of the semi-circle and the X-axis will change slightly as it passes across the star.
Because of how the method of deriving $\theta$ from $\phi$ is configured, we need this changing angle at every time-step instead of using the original 'static' $\phi$. The change in angle caused by this movement will be labelled $\psi$ and should be added to the original $\phi$ to create an 'updated' $\phi_{new}$.
To derive $\psi$ and make it clearer, $b$ has been re-labelled to $y$, where $x$ and $y$ are the coordinates of this 2D system with the center of the star at (0,0).} By differentiating,
\begin{eqnarray}
\textnormal{d}y = -r \cos{(\varpi+f)}\cos{i}\ \textnormal{d}f.
\label{eq:b}
\end{eqnarray}
Also explained in \cite{seager}, the x-coordinate of the center of the semi-circle will move as 
\begin{eqnarray}
x = -r\cos{(\varpi+f)},
\label{eq:x}
\end{eqnarray}
and so 
\begin{eqnarray}
\textnormal{d}x = -r \sin{(\varpi+f)}\ \textnormal{d}f.
\end{eqnarray}
Therefore $\psi$ can be calculated from
\begin{eqnarray}
\tan{\psi} =& \frac{\textnormal{d}y}{\textnormal{d}x},\\
\psi =& \arctan{[\cot{(\varpi+f)}\cos{i}]}.
\end{eqnarray}\\
Then, $\phi$ should be adjusted so that 
\begin{eqnarray}
\phi_{new} = \phi_{old} + \psi
\end{eqnarray}

\software{NumPy \citep{numpy}, Scipy \citep{scipy}, Matplotlib \citep{matplotlib}, batman \citep{Kreidberg:2015}, juliet \citep{juliet}, catwoman \citep{jones}.}




%% \section{Appendix information}


\bibliography{references}{}

\begin{thebibliography}{}
\expandafter\ifx\csname natexlab\endcsname\relax\def\natexlab#1{#1}\fi
\providecommand{\url}[1]{\href{#1}{#1}}
\providecommand{\dodoi}[1]{doi:~\href{http://doi.org/#1}{\nolinkurl{#1}}}
\providecommand{\doeprint}[1]{\href{http://ascl.net/#1}{\nolinkurl{http://ascl.net/#1}}}
\providecommand{\doarXiv}[1]{\href{https://arxiv.org/abs/#1}{\nolinkurl{https://arxiv.org/abs/#1}}}

\bibitem[{{Ackerman} \& {Marley}(2001)}]{AM}
{Ackerman}, A.~S., \& {Marley}, M.~S. 2001, \apj, 556, 872,
  \dodoi{10.1086/321540}

\bibitem[{{Ahlers} {et~al.}(2020){Ahlers}, {Johnson}, {Stassun}, {Col{\'o}n},
  {Barnes}, {Stevens}, {Beatty}, {Gaudi}, {Collins}, {Rodriguez}, {Ricker},
  {Vanderspek}, {Latham}, {Seager}, {Winn}, {Jenkins}, {Caldwell}, {Goeke},
  {Osborn}, {Paegert}, {Rowden}, \& {Tenenbaum}}]{ahlers}
{Ahlers}, J.~P., {Johnson}, M.~C., {Stassun}, K.~G., {et~al.} 2020, \aj, 160,
  4, \dodoi{10.3847/1538-3881/ab8fa3}

\bibitem[{{Alam} {et~al.}(2021){Alam}, {L{\'o}pez-Morales}, {MacDonald},
  {Nikolov}, {Kirk}, {Goyal}, {Sing}, {Wakeford}, {Rathcke}, {Deming},
  {Sanz-Forcada}, {Lewis}, {Barstow}, {Mikal-Evans}, \& {Buchhave}}]{alam}
{Alam}, M.~K., {L{\'o}pez-Morales}, M., {MacDonald}, R.~J., {et~al.} 2021,
  \apjl, 906, L10, \dodoi{10.3847/2041-8213/abd18e}

\bibitem[{{Batalha} {et~al.}(2017){Batalha}, {Mandell}, {Pontoppidan},
  {Stevenson}, {Lewis}, {Kalirai}, {Earl}, {Greene}, {Albert}, \&
  {Nielsen}}]{pandexo}
{Batalha}, N.~E., {Mandell}, A., {Pontoppidan}, K., {et~al.} 2017, \pasp, 129,
  064501, \dodoi{10.1088/1538-3873/aa65b0}

\bibitem[{Bourque {et~al.}(2021)Bourque, Espinoza, Filippazzo, Fix, King,
  Martlin, Medina, Batalha, Fox, Fowler, Fraine, Hill, Lewis, Stevenson,
  Valenti, \& Wakeford}]{exoctk}
Bourque, M., Espinoza, N., Filippazzo, J., {et~al.} 2021, The Exoplanet
  Characterization Toolkit (ExoCTK), 1.0.0,  Zenodo,
  \dodoi{10.5281/zenodo.4556063}

\bibitem[{{Buchner} {et~al.}(2014){Buchner}, {Georgakakis}, {Nandra}, {Hsu},
  {Rangel}, {Brightman}, {Merloni}, {Salvato}, {Donley}, \&
  {Kocevski}}]{PyMultiNest}
{Buchner}, J., {Georgakakis}, A., {Nandra}, K., {et~al.} 2014, \aap, 564, A125,
  \dodoi{10.1051/0004-6361/201322971}

\bibitem[{{Burrows} {et~al.}(2003){Burrows}, {Sudarsky}, \&
  {Hubbard}}]{BSH:2003}
{Burrows}, A., {Sudarsky}, D., \& {Hubbard}, W.~B. 2003, \apj, 594, 545,
  \dodoi{10.1086/376897}

\bibitem[{{Charbonneau} {et~al.}(2000){Charbonneau}, {Brown}, {Latham}, \&
  {Mayor}}]{c:2000}
{Charbonneau}, D., {Brown}, T.~M., {Latham}, D.~W., \& {Mayor}, M. 2000, \apjl,
  529, L45, \dodoi{10.1086/312457}

\bibitem[{{Dobbs-Dixon} {et~al.}(2012){Dobbs-Dixon}, {Agol}, \&
  {Burrows}}]{dobbs:2012}
{Dobbs-Dixon}, I., {Agol}, E., \& {Burrows}, A. 2012, \apj, 751, 87,
  \dodoi{10.1088/0004-637X/751/2/87}

\bibitem[{{Espinoza} {et~al.}(2017){Espinoza}, {Fortney}, {Miguel},
  {Thorngren}, \& {Murray-Clay}}]{espinoza}
{Espinoza}, N., {Fortney}, J.~J., {Miguel}, Y., {Thorngren}, D., \&
  {Murray-Clay}, R. 2017, \apjl, 838, L9, \dodoi{10.3847/2041-8213/aa65ca}

\bibitem[{{Espinoza} {et~al.}(2019){Espinoza}, {Kossakowski}, \&
  {Brahm}}]{juliet}
{Espinoza}, N., {Kossakowski}, D., \& {Brahm}, R. 2019, \mnras, 490, 2262,
  \dodoi{10.1093/mnras/stz2688}

\bibitem[{{Feroz} {et~al.}(2009){Feroz}, {Hobson}, \& {Bridges}}]{MultiNest}
{Feroz}, F., {Hobson}, M.~P., \& {Bridges}, M. 2009, \mnras, 398, 1601,
  \dodoi{10.1111/j.1365-2966.2009.14548.x}

\bibitem[{{Fortney}(2005)}]{fortney:2005}
{Fortney}, J.~J. 2005, \mnras, 364, 649,
  \dodoi{10.1111/j.1365-2966.2005.09587.x}

\bibitem[{{Fortney} {et~al.}(2010){Fortney}, {Shabram}, {Showman}, {Lian},
  {Freedman}, {Marley}, \& {Lewis}}]{fortney:2010}
{Fortney}, J.~J., {Shabram}, M., {Showman}, A.~P., {et~al.} 2010, \apj, 709,
  1396, \dodoi{10.1088/0004-637X/709/2/1396}

\bibitem[{{Greene} {et~al.}(2016){Greene}, {Line}, {Montero}, {Fortney},
  {Lustig-Yaeger}, \& {Luther}}]{greene:2016}
{Greene}, T.~P., {Line}, M.~R., {Montero}, C., {et~al.} 2016, \apj, 817, 17,
  \dodoi{10.3847/0004-637X/817/1/17}

\bibitem[{Harris {et~al.}(2020)Harris, Millman, van~der Walt, Gommers,
  Virtanen, Cournapeau, Wieser, Taylor, Berg, Smith, Kern, Picus, Hoyer, van
  Kerkwijk, Brett, Haldane, del R{'{\i}}o, Wiebe, Peterson,
  G{'{e}}rard-Marchant, Sheppard, Reddy, Weckesser, Abbasi, Gohlke, \&
  Oliphant}]{numpy}
Harris, C.~R., Millman, K.~J., van~der Walt, S.~J., {et~al.} 2020, Nature, 585,
  357, \dodoi{10.1038/s41586-020-2649-2}

\bibitem[{{Helling} {et~al.}(2020){Helling}, {Kawashima}, {Graham}, {Samra},
  {Chubb}, {Min}, {Waters}, \& {Parmentier}}]{helling:2020}
{Helling}, C., {Kawashima}, Y., {Graham}, V., {et~al.} 2020, \aap, 641, A178,
  \dodoi{10.1051/0004-6361/202037633}

\bibitem[{{Henry} {et~al.}(2000){Henry}, {Marcy}, {Butler}, \& {Vogt}}]{h:2000}
{Henry}, G.~W., {Marcy}, G.~W., {Butler}, R.~P., \& {Vogt}, S.~S. 2000, \apjl,
  529, L41, \dodoi{10.1086/312458}

\bibitem[{{Hubbard} {et~al.}(2001){Hubbard}, {Fortney}, {Lunine}, {Burrows},
  {Sudarsky}, \& {Pinto}}]{hubbard:2001}
{Hubbard}, W.~B., {Fortney}, J.~J., {Lunine}, J.~I., {et~al.} 2001, \apj, 560,
  413, \dodoi{10.1086/322490}

\bibitem[{Hunter(2007)}]{matplotlib}
Hunter, J.~D. 2007, Computing in Science \& Engineering, 9, 90,
  \dodoi{10.1109/MCSE.2007.55}

\bibitem[{Jones \& Espinoza(2020)}]{jones}
Jones, K., \& Espinoza, N. 2020, Journal of Open Source Software, 5, 2382,
  \dodoi{10.21105/joss.02382}

\bibitem[{{Kempton} {et~al.}(2017){Kempton}, {Bean}, \&
  {Parmentier}}]{kempton:2017}
{Kempton}, E. M.~R., {Bean}, J.~L., \& {Parmentier}, V. 2017, \apjl, 845, L20,
  \dodoi{10.3847/2041-8213/aa84ac}

\bibitem[{{Kipping}(2013)}]{kipping:2013}
{Kipping}, D.~M. 2013, \mnras, 435, 2152, \dodoi{10.1093/mnras/stt1435}

\bibitem[{{Kreidberg}(2015)}]{batman}
{Kreidberg}, L. 2015, \pasp, 127, 1161, \dodoi{10.1086/683602}

\bibitem[{{Kreidberg}(2018)}]{kreidberg:2018}
---. 2018, {Exoplanet Atmosphere Measurements from Transmission Spectroscopy
  and Other Planet Star Combined Light Observations}, 100,
  \dodoi{10.1007/978-3-319-55333-7_100}

\bibitem[{{Kreidberg} {et~al.}(2015){Kreidberg}, {Line}, {Bean}, {Stevenson},
  {D{\'e}sert}, {Madhusudhan}, {Fortney}, {Barstow}, {Henry}, {Williamson}, \&
  {Showman}}]{Kreidberg:2015}
{Kreidberg}, L., {Line}, M.~R., {Bean}, J.~L., {et~al.} 2015, \apj, 814, 66,
  \dodoi{10.1088/0004-637X/814/1/66}

\bibitem[{{Lewis} {et~al.}(2020){Lewis}, {Wakeford}, {MacDonald}, {Goyal},
  {Sing}, {Barstow}, {Powell}, {Kataria}, {Mishra}, {Marley}, {Batalha},
  {Moses}, {Gao}, {Wilson}, {Chubb}, {Mikal-Evans}, {Nikolov}, {Pirzkal},
  {Spake}, {Stevenson}, {Valenti}, \& {Zhang}}]{l2020}
{Lewis}, N.~K., {Wakeford}, H.~R., {MacDonald}, R.~J., {et~al.} 2020, \apjl,
  902, L19, \dodoi{10.3847/2041-8213/abb77f}

\bibitem[{{Line} \& {Parmentier}(2016)}]{LP:2016}
{Line}, M.~R., \& {Parmentier}, V. 2016, \apj, 820, 78,
  \dodoi{10.3847/0004-637X/820/1/78}

\bibitem[{{Line} {et~al.}(2013){Line}, {Wolf}, {Zhang}, {Knutson}, {Kammer},
  {Ellison}, {Deroo}, {Crisp}, \& {Yung}}]{chimera}
{Line}, M.~R., {Wolf}, A.~S., {Zhang}, X., {et~al.} 2013, \apj, 775, 137,
  \dodoi{10.1088/0004-637X/775/2/137}

\bibitem[{{MacDonald} {et~al.}(2020){MacDonald}, {Goyal}, \&
  {Lewis}}]{macdonald:2020}
{MacDonald}, R.~J., {Goyal}, J.~M., \& {Lewis}, N.~K. 2020, \apjl, 893, L43,
  \dodoi{10.3847/2041-8213/ab8238}

\bibitem[{{Mai} \& {Line}(2019)}]{chimera2}
{Mai}, C., \& {Line}, M.~R. 2019, \apj, 883, 144,
  \dodoi{10.3847/1538-4357/ab3e6d}

\bibitem[{{Mandel} \& {Agol}(2002)}]{ma:2002}
{Mandel}, K., \& {Agol}, E. 2002, \apjl, 580, L171, \dodoi{10.1086/345520}

\bibitem[{{Mordasini} {et~al.}(2016){Mordasini}, {van Boekel}, {Molli{\`e}re},
  {Henning}, \& {Benneke}}]{mordasini}
{Mordasini}, C., {van Boekel}, R., {Molli{\`e}re}, P., {Henning}, T., \&
  {Benneke}, B. 2016, \apj, 832, 41, \dodoi{10.3847/0004-637X/832/1/41}

\bibitem[{{Nikolov} \& {Sainsbury-Martinez}(2015)}]{n15}
{Nikolov}, N., \& {Sainsbury-Martinez}, F. 2015, \apj, 808, 57,
  \dodoi{10.1088/0004-637X/808/1/57}

\bibitem[{{{\"O}berg} {et~al.}(2011){{\"O}berg}, {Murray-Clay}, \&
  {Bergin}}]{oberg}
{{\"O}berg}, K.~I., {Murray-Clay}, R., \& {Bergin}, E.~A. 2011, \apjl, 743,
  L16, \dodoi{10.1088/2041-8205/743/1/L16}

\bibitem[{{Parmentier} {et~al.}(2020){Parmentier}, {Showman}, \&
  {Fortney}}]{p:2020}
{Parmentier}, V., {Showman}, A.~P., \& {Fortney}, J.~J. 2020, arXiv e-prints,
  arXiv:2010.06934.
\newblock \doarXiv{2010.06934}

\bibitem[{{Pluriel} {et~al.}(2020){Pluriel}, {Zingales}, {Leconte}, \&
  {Parmentier}}]{pluriel:2020}
{Pluriel}, W., {Zingales}, T., {Leconte}, J., \& {Parmentier}, V. 2020, \aap,
  636, A66, \dodoi{10.1051/0004-6361/202037678}

\bibitem[{{Pontoppidan} {et~al.}(2016){Pontoppidan}, {Pickering}, {Laidler},
  {Gilbert}, {Sontag}, {Slocum}, {Sienkiewicz}, {Hanley}, {Earl}, {Pueyo},
  {Ravindranath}, {Karakla}, {Robberto}, {Noriega-Crespo}, \&
  {Barker}}]{jwst-etc}
{Pontoppidan}, K.~M., {Pickering}, T.~E., {Laidler}, V.~G., {et~al.} 2016,
  Society of Photo-Optical Instrumentation Engineers (SPIE) Conference Series,
  Vol. 9910, {Pandeia: a multi-mission exposure time calculator for JWST and
  WFIRST}, 991016, \dodoi{10.1117/12.2231768}

\bibitem[{{Powell} {et~al.}(2019){Powell}, {Louden}, {Kreidberg}, {Zhang},
  {Gao}, \& {Parmentier}}]{powell:2019}
{Powell}, D., {Louden}, T., {Kreidberg}, L., {et~al.} 2019, arXiv e-prints,
  arXiv:1910.07527.
\newblock \doarXiv{1910.07527}

\bibitem[{{Rauscher} {et~al.}(2007){Rauscher}, {Menou}, {Seager}, {Deming},
  {Cho}, \& {Hansen}}]{r7}
{Rauscher}, E., {Menou}, K., {Seager}, S., {et~al.} 2007, \apj, 664, 1199,
  \dodoi{10.1086/519213}

\bibitem[{{Rein} \& {Ofir}(2019)}]{rein}
{Rein}, E., \& {Ofir}, A. 2019, \mnras, 490, 1111,
  \dodoi{10.1093/mnras/stz2556}

\bibitem[{{Ricker} {et~al.}(2014){Ricker}, {Winn}, {Vanderspek}, {Latham},
  {Bakos}, {Bean}, {Berta-Thompson}, {Brown}, {Buchhave}, {Butler}, {Butler},
  {Chaplin}, {Charbonneau}, {Christensen-Dalsgaard}, {Clampin}, {Deming},
  {Doty}, {De Lee}, {Dressing}, {Dunham}, {Endl}, {Fressin}, {Ge}, {Henning},
  {Holman}, {Howard}, {Ida}, {Jenkins}, {Jernigan}, {Johnson}, {Kaltenegger},
  {Kawai}, {Kjeldsen}, {Laughlin}, {Levine}, {Lin}, {Lissauer}, {MacQueen},
  {Marcy}, {McCullough}, {Morton}, {Narita}, {Paegert}, {Palle}, {Pepe},
  {Pepper}, {Quirrenbach}, {Rinehart}, {Sasselov}, {Sato}, {Seager},
  {Sozzetti}, {Stassun}, {Sullivan}, {Szentgyorgyi}, {Torres}, {Udry}, \&
  {Villasenor}}]{ricker-tess}
{Ricker}, G.~R., {Winn}, J.~N., {Vanderspek}, R., {et~al.} 2014, Society of
  Photo-Optical Instrumentation Engineers (SPIE) Conference Series, Vol. 9143,
  {Transiting Exoplanet Survey Satellite (TESS)}, 914320,
  \dodoi{10.1117/12.2063489}

\bibitem[{{Roman} \& {Rauscher}(2019)}]{rr:2019}
{Roman}, M., \& {Rauscher}, E. 2019, \apj, 872, 1,
  \dodoi{10.3847/1538-4357/aafdb5}

\bibitem[{{Sandford} \& {Kipping}(2019)}]{sandford:2019}
{Sandford}, E., \& {Kipping}, D. 2019, \aj, 157, 42,
  \dodoi{10.3847/1538-3881/aaf565}

\bibitem[{{Seager}(2010)}]{seager}
{Seager}, S. 2010, {Exoplanets}

\bibitem[{{Seager} \& {Sasselov}(2000)}]{ss:2000}
{Seager}, S., \& {Sasselov}, D.~D. 2000, \apj, 537, 916, \dodoi{10.1086/309088}

\bibitem[{{Sheppard} {et~al.}(2021){Sheppard}, {Welbanks}, {Mandell},
  {Madhusudhan}, {Nikolov}, {Deming}, {Henry}, {Williamson}, {Sing},
  {L{\'o}pez-Morales}, {Ih}, {Sanz-Forcada}, {Lavvas}, {Ballester}, {Evans},
  {Garc{\'\i}a Mu{\~n}oz}, \& {dos Santos}}]{s2020}
{Sheppard}, K.~B., {Welbanks}, L., {Mandell}, A.~M., {et~al.} 2021, \aj, 161,
  51, \dodoi{10.3847/1538-3881/abc8f4}

\bibitem[{{Sing} {et~al.}(2016){Sing}, {Fortney}, {Nikolov}, {Wakeford},
  {Kataria}, {Evans}, {Aigrain}, {Ballester}, {Burrows}, {Deming},
  {D{\'e}sert}, {Gibson}, {Henry}, {Huitson}, {Knutson}, {Lecavelier Des
  Etangs}, {Pont}, {Showman}, {Vidal-Madjar}, {Williamson}, \& {Wilson}}]{sing}
{Sing}, D.~K., {Fortney}, J.~J., {Nikolov}, N., {et~al.} 2016, \nat, 529, 59,
  \dodoi{10.1038/nature16068}

\bibitem[{Virtanen {et~al.}(2020)Virtanen, Gommers, Oliphant, Haberland, Reddy,
  Cournapeau, Burovski, Peterson, {Weckesser}, {Bright}, {van der Walt},
  {Brett}, {Wilson}, {Jarrod Millman}, {Mayorov}, {Nelson}, {Jones}, {Kern},
  {Larson}, {Carey}, {Polat}, {Feng}, {Moore}, {Vand erPlas}, {Laxalde},
  {Perktold}, {Cimrman}, {Henriksen}, {Quintero}, {Harris}, {Archibald},
  {Ribeiro}, {Pedregosa}, {van Mulbregt}, \& {Contributors}}]{scipy}
Virtanen, P., Gommers, R., Oliphant, T.~E., {et~al.} 2020, Nature Methods

\bibitem[{{von Paris} {et~al.}(2016){von Paris}, {Gratier}, {Bord{\'e}},
  {Leconte}, \& {Selsis}}]{vParis}
{von Paris}, P., {Gratier}, P., {Bord{\'e}}, P., {Leconte}, J., \& {Selsis}, F.
  2016, \aap, 589, A52, \dodoi{10.1051/0004-6361/201527894}

\bibitem[{{Wakeford} {et~al.}(2020){Wakeford}, {Sing}, {Stevenson}, {Lewis},
  {Pirzkal}, {Wilson}, {Goyal}, {Kataria}, {Mikal-Evans}, {Nikolov}, \&
  {Spake}}]{w2020}
{Wakeford}, H.~R., {Sing}, D.~K., {Stevenson}, K.~B., {et~al.} 2020, \aj, 159,
  204, \dodoi{10.3847/1538-3881/ab7b78}

\bibitem[{{Welbanks} {et~al.}(2019){Welbanks}, {Madhusudhan}, {Allard},
  {Hubeny}, {Spiegelman}, \& {Leininger}}]{wellbanks}
{Welbanks}, L., {Madhusudhan}, N., {Allard}, N.~F., {et~al.} 2019, \apjl, 887,
  L20, \dodoi{10.3847/2041-8213/ab5a89}

\bibitem[{{Williams} {et~al.}(2006){Williams}, {Charbonneau}, {Cooper},
  {Showman}, \& {Fortney}}]{w6}
{Williams}, P. K.~G., {Charbonneau}, D., {Cooper}, C.~S., {Showman}, A.~P., \&
  {Fortney}, J.~J. 2006, \apj, 649, 1020, \dodoi{10.1086/506468}

\end{thebibliography}
\bibliographystyle{aasjournal}



\end{document}